\documentclass[letterpaper]{article} % Aptara syntax

\usepackage[ruled,vlined]{algorithm2e}

\usepackage{helvet}
\usepackage{courier}
\usepackage{mathrsfs}
\usepackage{graphicx}
\usepackage{amsmath}
\usepackage{amssymb}
\usepackage{amsthm}
\usepackage{multirow} 
\usepackage{subcaption}
\usepackage{tikz}
\usetikzlibrary{arrows, petri,topaths}
\usepackage{tkz-berge}
\newtheorem{theorem}{Theorem}
\newtheorem{corollary}[theorem]{Corollary}
\newtheorem{proposition}[theorem]{Proposition}

\newtheorem{lemma}[theorem]{Lemma}

\newtheorem{observation}[theorem]{Observation}

\theoremstyle{definition}

\newtheorem{definition}{Definition}

\DeclareMathOperator{\argmin}{argmin}
\DeclareMathOperator{\argmax}{argmax}

\newenvironment{rlemma}[1]{\medskip\noindent\textbf{Lemma~\ref{#1}.}\begin{itshape}}{\end{itshape}}
\newenvironment{rproposition}[1]{\medskip\noindent\textbf{Proposition~\ref{#1}.}\begin{itshape}}{\end{itshape}}

\newcommand{\eps}{\epsilon}
\renewcommand{\vec}[1]{\mathbf{#1}}

\newcommand{\coursename}{(67686) Mathematical Foundations of AI}

\newcommand{\handout}[5]{
   \renewcommand{\thepage}{#1-\arabic{page}}
   \noindent
   \begin{center}
   \framebox{
      \vbox{
    \hbox to 5.78in { {\bf \coursename}
         \hfill #2 }
       \vspace{4mm}
       \hbox to 5.78in { {\Large \hfill #5  \hfill} }
       \vspace{2mm}
       \hbox to 5.78in { {\it #3 \hfill #4} }
      }
   }
   \end{center}
   \vspace*{4mm}
}

\def\ol{\overline}

\newcommand{\ceil}[1]{\left\lceil #1 \right\rceil}
\newcommand{\floor}[1]{\left\lfloor #1 \right\rfloor}

\newenvironment{proof-sketch}{\noindent{\bf Sketch of Proof}\hspace*{1em}}{\qed\bigskip}
\newenvironment{proof-idea}{\noindent{\bf Proof Idea}\hspace*{1em}}{\qed\bigskip}
\newenvironment{proof-of-lemma}[1]{\noindent{\bf Proof of Lemma #1}\hspace*{1em}}{\qed\bigskip}
\newenvironment{proof-attempt}{\noindent{\bf Proof Attempt}\hspace*{1em}}{\qed\bigskip}

\makeatletter
\def\fnum@figure{{\bf Figure \thefigure}}
\def\fnum@table{{\bf Table \thetable}}
\long\def\@mycaption#1[#2]#3{\addcontentsline{\csname
  ext@#1\endcsname}{#1}{\protect\numberline{\csname
  the#1\endcsname}{\ignorespaces #2}}\par
  \begingroup
    \@parboxrestore
    \small
    \@makecaption{\csname fnum@#1\endcsname}{\ignorespaces #3}\par
  \endgroup}
\def\mycaption{\refstepcounter\@captype \@dblarg{\@mycaption\@captype}}
\makeatother

\newcommand{\mathify}[1]{\ifmmode{#1}\else\mbox{$#1$}\fi}
\newcommand{\bigO}O

\newcommand\tup[1]{\left\langle #1 \right\rangle}

\newcommand{\QQ}{{\cal Q}}

\renewcommand{\vec}[1]{{\mathbf #1}}

\def\VV{{\cal V}}

\newcommand{\remove}[1]{{}}

\def\SS{{\cal S}}

\newcommand{\xqed}{\mbox{\raggedright $\Diamond$}}

\newcommand{\step}[1]{\stackrel{{\scriptscriptstyle{#1}}}{\rightarrow}}

\newcommand{\newpar}[1]{
\vspace{-1mm}
\paragraph{#1}}
\newcommand{\newsubsec}[1]{
\vspace{-1.5mm}
\subsection{#1}}

\newcommand{\omittext}[1]{}

\def\shortcite{\cite}

\newcount\Comments % 0 suppresses notes to selves in text
\definecolor{darkgreen}{rgb}{0,0.6,0}
\newcommand{\kibitz}[2]{\ifnum\Comments=1{\color{#1}{#2}}\fi}
\newcommand{\rmr}[1]{\kibitz{blue}{[RESHEF:#1]}}

\begin{document}

\title{Plurality Voting under Uncertainty}
\author{Reshef Meir}
\maketitle

\begin{abstract}
Understanding the nature of strategic voting is the holy grail of social choice theory, where game-theory, social science and recently  computational approaches are all applied in order to model the incentives and behavior of voters.

In a recent paper, Meir et al.~\shortcite{MLR14} made another step in this direction, by suggesting a behavioral game-theoretic model for voters under uncertainty. For a specific variation of best-response heuristics, they proved initial existence and convergence results in the Plurality voting system.

%, where voters observe candidates' prospective scores (e.g., the results of a poll or a previous round), and believe that the true scores are within some range of those. Voters  refrain from actions that are \emph{locally-dominated}, i.e., dominated according to their beliefs. %Meir et al. show that when voters with the same beliefs start from the truthful profile and may change their vote in turns,  the game must converge to an equilibrium. 
In this paper, we extend the model in multiple directions, considering voters with different uncertainty levels, simultaneous strategic decisions, and a more permissive notion of best-response. We prove that a voting equilibrium exists even in the most general case. Further, any society voting in an iterative setting is guaranteed to converge.

We also analyze an alternative behavior where voters try to minimize their worst-case regret. We show that the two behaviors coincide in the simple setting of Meir et al., but not in the general case. 
\end{abstract}

\section{Introduction}
Suppose that your favorite candidate in the elections trails behind in the polls. A game-theorist advice in such a case would be to ``vote strategically.'' 

%Game-theoretic models of voting have been studied for decades now, yet there is no consensus over a single model that describes the behavior of strategic voters. 

As a simple example, consider an election with 3 candidates $M=\{a,b,c\}$. Suppose that the most recent poll indicates that $a$ is leading with $45\%$ of the votes,  whereas there are $40\%$  for $b$ and $15\%$ for $c$ (see Fig.~\ref{fig:ex}). 
A voter who prefers $c$ over $b$ over $a$ may conclude that $c$ has no chance of winning, and thus a wise strategic decision would be to vote for $b$.  However, there is no consensus whatsoever regarding how to generalize this observation to an arbitrary situation, or even on the conditions under which $c$ should be deserted.

This lack of a conclusive notion for strategic voting is partly due to the fact that there are many ways to describe the information voters have, as well as their beliefs over other voters' preferences and actions when casting their vote. Moreover, even for a given belief there may be several different actions that can be justified as rational. Hence any model of voting behavior and equilibrium must state explicitly its epistemic and behavioral assumptions.

We can of course apply some ``standard'' assumptions from game-theory, for example that voters play a Nash equilibrium of the game induced by their preferences. However, such a prediction turns out to be very uninformative: a single voter can rarely affect the outcome, and thus almost any way the voters vote is a Nash equilibrium (including, for example, when all voters vote for their least preferred candidate). 

It is therefore natural to consider the role of \emph{uncertainty} in voters' decisions. One of the most prominent models for voting under uncertainty was suggested by Myerson and Weber~\shortcite{MW93}, where voters preferences are sampled from a known prior distribution. An equilibrium according to the MW model is a mapping from preferences to a distribution over votes, such that each voter maximizes her expected utility w.r.t. this distribution. Myerson and Weber prove via a fixed-point argument that an equilibrium always exists for every positional scoring rule, and in particular for Plurality. 

The MW model reflects a classical economic approach (as do many other game-theoretic models of voting), but its epistemic and behavioral assumptions seem highly unreasonable in the context of voting.
As studies in behavioral psychology show, human decision makers often ignore probabilistic information even when it is given, employing various heuristics instead~\cite{tversky1974judgment}.
 It is thus unlikely that people  are  able to represent such complex distributions, or to compute and optimize their own expected utility, let alone the equilibria of the game. % Thus it is not clear how voters would reach an equilibrium, or even realize that they are playing one. 

In a recent effort to reconcile well-founded decision making approaches with a formal game-theoretic analysis, Meir et al.~\shortcite{MLR14} suggested a model for Plurality voting relying on \emph{strict uncertainty} and \emph{local dominance}. Informally, voters' beliefs can be described by a single vector of candidates' scores. This prospective score vector may be the result of a poll, derived from acquaintance with the other voters, from the outcome of a previous round of voting, etc. Each voter has a single uncertainty parameter, reflecting how sure she is in the correctness of the prospective scores---higher uncertainty means the voter considers a larger range of outcomes (score vectors) as ``possible''. In the example introduced above, a voter with an intermediate uncertainty parameter will consider a tie between $a$ and $b$ as ``possible,'' but will reject any outcome where $c$ can win as impossible.

Given this uncertain, non-probabilistic view, it is not a-priori clear how a voter should act.  Meir et al. adopted one of the classic approaches to decision making under strict uncertainty, assuming that a voter $i$ will refrain from voting for a candidate $a_i$ that is \emph{locally dominated} by another candidate $a'_i$~\cite{aumann1995backward,aumann1999interactive}. Intuitively, $a'_i$ \emph{locally-dominates} $a_i$ if it is always at least as good, and sometimes strictly better, to vote for $a'_i$ than for $a_i$ (taking into account all score vectors that voter $i$ considers possible).  A \emph{voting equilibrium} is simply a state where no voter votes for a locally-dominated candidate according to her beliefs. Going back to our example, our voter will consider candidate $c$ as locally-dominated by candidate $b$, and will thus strategize by voting for $b$. 

Informally, Meir et al.~\shortcite{MLR14} proved that if all voters have the \emph{same uncertainty parameter},  start from the truthful state, and play one at a time, then they always converge to a voting equilibrium. In the special case of zero uncertainty, this simply means that the best-response dynamics converges to a Nash equilibrium (which is known from \cite{MPRJ:2010:AAAI}).

\rmr{not here:Since a candidate may be dominated by several other candidates, Meir et al. made an additional assumption that the voter will select her most favorite candidate among those.}

%Under the same assumption they provided another convergence result, namely that from any initial state there is \emph{some path} (i.e., some order over voters) that converges to an equilibrium. Both convergence results hold under several common distance metrics.  
However, these assumptions are rather restrictive. Some voters may be less informed than others, or simply more ``stubborn'' (e.g., require even fewer votes to $c$ in order to be convinced it cannot win). Also, there is no reason to believe that voters initially vote for their most preferred candidate, or that only one voter changes her vote after each poll or voting round (see, e.g., \cite{reijngoud2012voter}). Meir et al.~\shortcite{MLR14} showed \emph{empirically} that convergence occurs even without those assumptions, and conjectured that at least some of the assumptions could be relaxed. However they provide no formal guarantee for such convergence,%\footnote{Meir et al.~\shortcite{MLR14} claim that from any initial state there is \emph{some} path that converges, but this is a weaker claim, and we also suspect that their proof is wrong. See Appendix~\ref{apx:flaw}.}  
and in fact for voters with different uncertainty levels even existence of equilibrium remains an open question.

\subsection{Research goals}
The main purpose of this work is to close this gap, and to prove equilibrium existence and convergence under conditions that are as broad as possible.

While we adopt the model of Meir et al. at large, we study a non-atomic variation of it, where the effect of any single voter is negligible (as in \cite{MW93}). This simplifies the model, and allows us to prove stronger results when arbitrary subsets of voters change their vote simultaneously. However all of our results hold in the finite case, for voters that move one-at-a-time (see Appendix~\ref{apx:finite}). We modify the distance function used to determine the possible ranges of candidates' scores, to one that is better justified by psychological studies (for details see Footnote~\ref{fn:KT}).

%We focus on a particular metric, where the uncertainty over the score of each candidate is proportional to its prospective score (the multiplicative metric). This means that the amount of uncertainty does not depend on the number of voters, which is a convenient assumption for infinite populations. More importantly, strong evidence from psychological experiments suggests that people treat uncertainty as scale-invariant~\cite{kahneman1974subjective} (for details see Footnote~\ref{fn:KT}).

%We also initiate a deeper study the local-dominance behavior.
In addition, we are interested  whether the behavioral assumptions can be weakened%(in the same sense that better-response is a weak variation of best-response)
, or, alternatively, be replaced with a more nuanced way  to select among several undominated candidates. To that end, we define a variation of the model where voters are aiming to minimize their \emph{worst-case regret} over all states they believe possible, and compare this behavior to voting under local-dominance.

\subsection{Our contribution}
Our main result is proving that voters with local-dominance behavior always converge to an equilibrium. This holds for any mixture of types (including voters with different uncertainty levels), for any initial voting profile, and for any order of moves (including moves of arbitrary subsets of voters, and suboptimal moves).

We then turn to study voters minimizing worst-case regret. When voters have the same uncertainty level and start from the truthful profile, we show that regret minimization coincides with local dominance.  However if these requirements are relaxed then the behaviors may significantly differ, and even the existence of equilibrium is not guaranteed.

%While we assume nonatomic voters in this paper (where every voter has a negligible impact), almost all of our results also hold for a finite set of voters that play iteratively one at a time.  Therefore our main result settles a conjecture from \cite{MLR14} about equilibrium existence for $n$ voters with different uncertainty levels.
	 
\section{The Formal Model}

\paragraph{Basic notations}
Where possible, we follow the notations and definitions of \cite{MLR14}. We make the necessary changes to adapt their model to nonatomic voters.
We denote $[x]=\{1,2,\ldots,x\}$ for all $x\in\mathbb N$. For a finite set $A$, $\Delta(A)$ denotes the set of all probability distributions over $A$ (all non-negative vectors of size $|A|$ that sum to $1$).

The set of candidates is denoted by $M$, where $m=|M|$.  
Let $\QQ=\pi(M)$ be the set of all strict orders (permutations) over $M$. We encode all the information on a voter, including her preferences, in her \emph{type},\footnote{We later extend the definition of a type to also include the belief structure and behavior of the voter.}  where the set of all types is denoted by $\VV$.

Thus a voter of type $v\in \VV$ has preferences $Q_v\in \QQ$, where $Q_v(a)\in[m]$ is the rank of candidate $a\in M$ (lower is better), and $q_v=Q^{-1}_v(1)$ is her most-preferred candidate.  We denote $a \succ_v b$ if $Q_v(a) < Q_v(b)$.

We do not have a finite set of voters. Rather, a \emph{preference profile} $\vec Q\in \Delta(\QQ)$ is a distribution over preferences, specifying the fraction of voters with each preference order. A \emph{population} $\vec V\in \Delta(\VV)$ is a distribution over \emph{types}, that is, a preference profile aggregated with any additional information specified by voters' types (e.g., their beliefs and behaviors).

%The Plurality voting rule $f$ allows voters to vote for a single candidate from the set $M$. Then, $f$ chooses the candidate with the highest score. 
%Thus any population defines a nonatomic \emph{game}, where voters' types determine their utilities.
Under the Plurality rule, every voter selects a single candidate. Intuitively, we should specify how many voters of each type vote for every candidate.
Formally, an \emph{action profile} (also called \emph{state}) $\vec a$ is a refinement of $\vec V$, where $a(v,c)\in \mathbb R_+$ denotes the fraction of voters of type $v$ who vote for $c\in M$. Our next definitions are intended to enable notations and analysis that are similar to those used in the atomic case. 

Since a single voter has negligible influence, we only consider moves by subsets of voters whose size is a multiple of $\eps$ (for some arbitrarily small $\eps$). All voters in each set have the same type and move simultaneously (although they need not be aware of this, see also Appendix~\ref{apx:acyclic}).
%We can thus number these sets and treat and arbitrary voter in the set as an identified voter.  
We denote by $I$ the collection of these $1/\eps$ sets. Since all voters in set $i\in I$ are indistinguishable we refer to ``voter $i$'' which is an arbitrary voter in the set $i$. This voter has a well-defined type $v_i$ and a well defined action $a_i$ in every action profile $\vec a$. 
%
%As the number of types is finite,  define $I = \{\tup{v,c}\}_{v\in \VV,c\in M}$. Each $\tup{v_i,a_i}\in I$ is a \emph{voter identifier}. Thus we can talk about an individual non-atomic voter $i$, which has a well-defined type $v_i$ and a well-defined action $a_i$.%\footnote{We use $c$ to denote a non-specific candidate, and $a_i$ to denote the candidate that voters of identifier $i$ vote for.} Let $w_i=a(v_i,c_i)$, and note that two voters with the same identifier $i$ are indistinguishable. 
%%, thus we only care about the total frequency $a(v_i,a_i)$ of each identifier.  
 %Thus we can rewrite $\vec a=(a(v,c))_{v\in \VV,c\in M}$ as $\vec a = (w_i)_{i\in I}$. Note that a voter's type is fixed, whereas her identifier depends on her action.
%%Then, to specify a particular action profile $\vec a$, we only need to specify the total amount $w_i$ of voters of each identifier $i$, that is, $\vec a = (w_i)_{i\in I}$. 
%Let $I(\vec a)$ be the \emph{support} of $\vec a$, i.e., $I(\vec a) = \{i\in I: w_i>0\}$. 

\newpar{Winner determination and tie-breaking}
The outcome of the Plurality rule in state $\vec a$, denoted $f(\vec a)\in M$, only depends on the total number of votes for each candidate. 
We thus define the \emph{score vector} $\vec s_{\vec a}$ induced by action profile $\vec a$. That is, $s_{\vec a}(c) = \sum_{v\in \VV}a(v,c)=|\{i\in I: a_i=c\}|\eps$.
We will use $\vec a$ and $\vec s_{\vec a}$ interchangeably, sometimes omitting the subscript $\vec a$. We denote by $\SS=\mathbb R_+^m$ the set of all score vectors. We also sometimes refer to score vectors as states, although $\vec s$ may not be attained from an actual action profile. E.g., it is possible that $\sum_{c\in M}s(c)\neq 1$. Note that we may only use $\vec s$ in a context where voters' identifiers are not important. 

 Note that changing the vote of a single voter in $\vec a$ does not change $\vec a$ or $\vec s_{\vec a}$, yet we need a way to settle tie-breaking. %, and there must be cases where a voter considers herself pivotal. % In what follows we will only consider moves by \emph{groups} of voters that have a non-zero weight (thus the scores actually change). 
Every score vector $\vec s$ implicitly contains an arbitrary ``tie-breaker'' $Q^+_{\vec s}\in \QQ$.  

The \emph{winner} $f(\vec s)$ is the candidate $c\in M$ whose score $s(c)$ is maximal. If there is more than one candidate with maximal score, we break ties according to $Q^+_{\vec s}$. Formally, $f(\vec s) = \argmin_{c\in\argmax s(c')}Q^+_{\vec s}(c)$. Note that the outcome is well defined even for score vectors that are not derived from valid action profiles.

\newpar{Dynamics and equilibria}
%The way voters vote depends not only on their preferences, but also on their beliefs. A standard game-theoretic assumption is that rational players aim at playing the optimal response to the actions of other players. However in most cases the action profile is not known exactly but only approximately. %Based on whatever information they have on the profile, voters can each decide on their action.
We describe the behavior of a voter of type $v\in\VV$ by a \emph{response function} $g_v:M \times \SS \rightarrow 2^M\setminus \emptyset$. That is, a mapping from the current state (only taking into account aggregate scores) and current action, to a subset of actions. 
Together, $Q_v$ and $g_v$ completely define the type $v$. For an identified voter $i\in I$, we write $g_i(\vec s)$ instead of $g_{v_i}(a_i,\vec s)$.  We also write $g_i(\vec a)$ as a shorthand for $g_i(\vec s_{\vec a})$. 
Intuitively, this means that a voter $i\in I$  may choose any action in $g_i(\vec a)$. %, where the way we construct $g_i$ should reflect whatever voter $i$ believes in state $\vec a$. %\footnote{We can also think of more general response functions, that depend on the history of the game. However in this paper the response only depends on the current state.} 
In the terms of \cite{jaggard2011distributed}, $g_i$ is a \emph{historyless}, non-deterministic response function.

 \begin{definition}
\label{def:eq_NA} % Let $\vec V$ be a population of voters. 
A \emph{voting equilibrium}  for population $\vec V$ is a state $\vec a$, where $g_i(\vec a) = \{a_i\}$ for all $i\in I$.
\end{definition}
For example, if $g_i$ is the \emph{best-response function} ($g_i(\vec a)=\{c\in M: c \text{ maximizes the utility of $i$ in $\vec a$}\}$ for all $i$), then a voting equilibria coincides with the Nash equilibria of the  game. We emphasize that classic results on existence of equilibria in nonatomic games (e.g., ~\cite{schmeidler1973equilibrium}) do not apply, since even if we assign cardinal utilities to voters, those would be highly discontinuous in the action profile. \rmr{Simon and Zame 1990?} 
Note that as in \cite{MW93}, % but in contrast with some other previous works~\cite{?},
 we do not assume that voters with the same type, or even the same identifier, are coordinated.

Even in cases where an equilibrium exists, players may or may not reach one, depending on their initial state and the order in which they play. We are therefore interested in sufficient conditions under which the game is \emph{acyclic}, i.e. there are no finite cycles of rational moves by groups of any size.\footnote{In finite games, acyclicity and guaranteed convergence imply one another. In nonatomic games the relation is a bit more delicate, see Appendix~\ref{apx:acyclic}.} 

\section{Uncertainty and Strategic Voting}
The most important part of the model is of course the way we define the behavior, which is the response function $g_i$. 
This behavior depends on voters' preferences, and also on their beliefs about the current state. We assume voters derive their beliefs using a distance-based strict uncertainty model, following \cite{MLR14}.

%\documentclass[]{article}
%\usepackage{tikz}
%\usepackage{subcaption}
%%\usepackage{tikz}%,fullpage}
%\usetikzlibrary{arrows,%
                %petri,%
                %topaths}%
%\usepackage{tkz-berge}
%
%\begin{document}
%
%%%%%%%%%%%%%%%%%%%

\begin{figure}
\centering

\begin{tikzpicture}[scale=1,transform shape]

\tikzstyle{vote1}=[draw=black,fill=white,
inner sep=0pt,minimum size=5mm]
\tikzstyle{vote2}=[draw=black,fill=black!10!white,
inner sep=0pt,minimum size=5mm]
\tikzstyle{vote3}=[draw=black,fill=black!30!white,
inner sep=0pt,minimum size=5mm]
\tikzstyle{vote4}=[draw=black,fill=black!50!white,
inner sep=0pt,minimum size=5mm]
\tikzstyle{vote5}=[draw=black,fill=black!70!white,text=white,
inner sep=0pt,minimum size=5mm]

	\draw[thick] (0.5,0.2) -- (3.5,0.2);
	\node at (1,0) {$a$};
	\node at (2,0) {$b$};
	\node at (3,0) {$c$};
	%\draw[dashed] (0,1.5) -- (6,1.5);
	
	\draw (0.75,0.3) rectangle (1.25,4.5);
	\node at (1.2,4.9) {$s(a)=0.45$};
	
	\draw (1.75,0.3) rectangle (2.25,4);
	\node at (2.2,4.3) {$s(b)=0.4$};
	
	\draw (2.75,0.3) rectangle (3.25,1.5);
	\node at (3.2,1.8) {$s(c)=0.15$};

	\draw[thick] (4.5,0.2) -- (7.5,0.2);
	\node at (5,0) {$a$};
	\node at (6,0) {$b$};
	\node at (7,0) {$c$};
	
	\draw[dashed] (4.5,3.4) -- (7.5,3.4);
	
	\draw (4.75,0.3) rectangle (5.25,4.5);
	\draw[|<->|,dashed,thick] (5,3.91) -- (5,5.2);

	\draw (5.75,0.3) rectangle (6.25,4);
	\draw[|<->|,dashed,thick] (6,3.47) -- (6,4.6);

	\draw (6.75,0.3) rectangle (7.25,1.5);
	\draw[|<->|,dashed,thick] (7,1.75) -- (7,1.3);

\end{tikzpicture}

\caption{\label{fig:ex}On the left figure we see a given state $\vec s$, which can be thought of as the result of a poll, or of the current voting round. On the right we depict the set of possible states $S(\vec s,r)$ for $r = 0.15$. Any $\vec s'$ is a possible state as long as the score of each candidate is in the marked range. The dashed line marks the threshold above which a candidate is considered a possible winner.}
\end{figure}
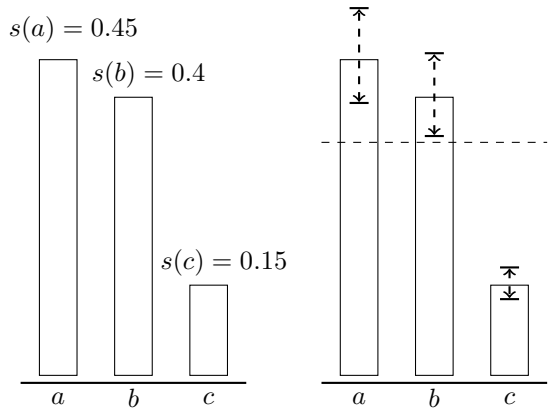
%\end{document}
\subsection{Distance-based uncertainty}
Suppose we have some distance measure for score vectors, denoted by $\delta(\vec s,\vec s')$.
For any $\vec s$, let $S(\vec s,x)\subseteq \SS$ be  the set of vectors that are at distance at most $x$ from $\vec s$. Formally, $S(\vec s,x)=\{\vec s' : \delta(\vec s,\vec s') \leq x\}$. 
A distance function $\delta$ is \emph{candidate-wise} if it can be written as $\delta(\vec s,\vec s') = \max_{c\in M} \delta'(s_c,s'_c)$ for some monotone function $\delta'$ (meaning that for a fixed $s$, $\delta'(s,s')$ is nondecreasing in $|s-s'|$). Thus $\vec s'\in S(\vec s,x)$ only if the score of every candidate in $\vec s'$ is sufficiently close to its score in $\vec s$, see Fig.~\ref{fig:ex}.

In \cite{MLR14}, several metrics have been suggested, including various $\ell_d$ norms and the earth-mover distance. %\footnote{We use the term `distance' but make no requirement of symmetry or triangle inequality.}   
 Among those the multiplicative distance (where $\delta'(s,s')=\max\{s'/s, s/s'\}-1$) and the $\ell_\infty$ metric (where $\delta'(s,s') = |s-s'|$) are candidate-wise.  

While all of our results apply for any candidate-wise distance, for concreteness we assume throughout the paper the multiplicative distance. Note that the multiplicative distance is independent of the amount of voters, i.e. $\delta(\vec s,\vec s') = \delta(\alpha \vec s,\alpha \vec s')$ for all $\alpha>0$. This is consistent with findings on how people perceive uncertainty over numerical values~\cite{kahneman1974subjective,tversky1974judgment}.\footnote{\label{fn:KT}For example, the studies show that if the average number of girls born daily in a hospital is $s$, then people believe that the probability that on a given day the number is within $[(1-r)s,(1+r)s]$ is fixed and does not depend on $s$. Kahneman and Tversky highlight that this reasoning stands in contrast to the  scientific truth in this case, where the range $r$ is proportional to $1/\sqrt s$. In our case we can think of the score of a candidate as the limit of a Poisson variable (rather than Binomial in the hospital example), and thus the range $[s/(1+r),s(1+r)]$ is more appropriate.}
%According to the \emph{multiplicative distance}, $\delta(\vec s,\vec s' ) \leq x$ if for all $c\in M$, both $s'(c) \leq s(c)(1+x)$ and $s(c) \leq s'(c)(1+x)$. Intuitively, this means that the score of each candidate can change (either increase or decrease) by a factor of at most $(1+x)$.
%In this paper we assume that $\delta$ is the multiplicative distance unless explicitly stated otherwise.\footnote{All of our results hold for the $\ell_\infty$ metric as well, where $\delta(\vec s,\vec s') = \max_{c\in M}|s'(c)-s(c)|$.} 

We define the \emph{uncertainty parameter} $r_v\in \mathbb R_+$ as part the agent's type (will be later used to construct the response function  $g_v$). Given a profile $\vec a$%(an estimate of the outcome based on polls, acquaintance with the other voters, previous voting rounds, etc.)
, voter $i\in I$ believes that the actual state may be any $\vec s' \in S(\vec s_{\vec a},r_i)$, where $r_i = r_{v_i}$. 

Note that the distance between two score vectors that only differ by the tie-breaker is $0$.
Thus if several candidates $C\subseteq M$ are tied in $\vec s=\vec s_{\vec a}$ with maximal score, for each $c\in C$ there is a state $\vec s' \in S(\vec s,r_i)$ where $f(\vec s')=c$ (any state where $Q^+_{\vec s'}(c)=1$), and this holds for any $r_i$.

%Note that in contrast to the atomic case~\cite{MLR14}, we get a simpler definition of accessible states $S$ that does not depend on the identity of the voter. 
%
%
%
%The $\delta$ distance may be a $\ell_d$ norm for some $d\geq 1$. Thus $\delta_{\ell_1}(\vec s',\vec s) \leq x$ means that we can attain $\vec s'$ by adding/removing a total of $x$ voters to $\vec a$ (note that the total number of votes in $\vec s,\vec s'$ may be different). Similarly, the $\ell_\infty$ norm means that we can add or remove at most $x$ votes for each candidate.
%
%
%In this work we will focus on the $\ell_\infty$ distance, but all of our results apply for the multiplicative distance as well. 

A voter facing strict uncertainty may use various heuristics when selecting a strategy. In this work we follow two standard approaches: avoiding dominated strategies~\cite{aumann1999interactive}, and minimizing worst-case regret~\cite{savage1951theory,hyafil2004regret}.

\newpar{Voter influence}
 In the nonatomic case, every voter has negligible influence. However, this influence is not perceived by the voter as zero, as otherwise voters would always be indifferent about their actions. We define $f(\vec s,c)$ to be the outcome that a voter expects under score vector $\vec s$, when voting for $c$. Intuitively,  the extra vote $c$ decides the winner if several candidates are tied with maximal score in $\vec s$ (overriding the default tie-breaker $Q^+_{\vec s}$), and otherwise has no effect. A more formal definition appears in Appendix~\ref{apx:influence}.

\newsubsec{Local dominance}
The first approach we consider follows \cite{conitzer2011dominating,MLR14}, where voting dynamics is determined by local dominance relations.

Consider a particular voter of identifier $i=\tup{v_i,a_i}$. 
Let $S=S(\vec s,r_i)$. 
%\begin{definition}
%\label{def:beat}
We say that action $a_i$ \emph{$S$-beats} $b_i$  if there is at least one state $\vec s'\in S$ s.t.\ $f(\vec s',a_i) \succ_i f(\vec s', b_i)$. That is where $i$ strictly prefers $f(\vec s',a_i)$ over $f(\vec s',b_i)$.
%\end{definition}
%\begin{definition}[Local dominance]
%\label{def:local}
Action $a_i$ \emph{$S$-dominates} $b_i$ if (I) $a_i$ $S$-beats $b_i$; and (II) $b_i$ does \emph{not} $S$-beat $a_i$.
%\end{definition}

We next define the response function that strategic voters apply under the local-dominance behavioral model. 
%A step of a voter $i$ from candidate $a$ to another candidate $a'$ is called a \emph{strategic response}, and is denoted by $a \step{i} a'$. The following definition describes the basic strategic response in our model.
Consider a voter $i$, where $v_i=\tup{Q_i,r_i}$. 
\begin{definition}[Strategic move under local dominance (LD)]
\label{def:LD}
Let $a_i$ be the vote of $i$ in profile $\vec a$.   
Let $D\subseteq M$ be the set of candidates that $S(\vec a,r_i)$-dominate $a_i$, and that are not $S(\vec a,r_i)$-dominated. If $D=\emptyset$ then $g_i(\vec a)=\{a_i\}$. Otherwise:
\begin{itemize}
	\item For a \emph{weak LD voter}, $g_i(\vec a)= D$. 
	\item For a \emph{strict LD voter}, $g_i(\vec a) = \argmin_{d\in D} Q_i(d)$.
\end{itemize}
\end{definition}
In other words,  a strict LD voter votes  the most preferred  candidate that locally-dominates her current choice (if such exists), which is   almost identical to the definition in \cite{MLR14} for finite populations (see also Def.~\ref{def:LD_finite} in Appendix). The definition of a weak LD voter is much more permissive, and does not restrict the voter to select the most preferred candidate in $D$. 
A higher value of $r_i$ may either indicate that the voter is less informed, or simply that she requires stronger evidence that a move will be beneficial, a tendency we can interpret as stronger \emph{loss aversion}~\cite{kahneman1991anomalies}.

	 As in the model of \cite{MLR14}, we get the following immediate observation.

\begin{proposition}
\label{th:eq_dominated_NA}
Let $\vec V$ be a population of LD voters. A profile $\vec a$ is a voting equilibrium iff  $\forall i\in I, !\exists a'_i\in M,$ such that $a'_i$ $S(\vec a,r_i)$-dominates $a_i$. I.e., if no voter votes for a locally dominated candidate. 
\end{proposition}

%If a game is guaranteed to converge for \emph{some} scheduler, then it has a weak potential~\cite{?}.
	
\section{Convergence with LD Voters}
%
%
	%\rmr{needed?}
%Let $H_w(\vec s)\subseteq M$ be the set of candidates that need exactly $w$ more votes to become the winner.  Let $\ol H_w(\vec s) = \bigcup_{w'\leq w} H_{w'}(\vec s) = \{c : s(c) \geq s(f(\vec s))-w\}$. 

\newsubsec{Strategic responses and possible winners}

We say that candidate $c$ is a \emph{possible winner for $i$ in state $\vec s$} if there is a possible state where $c$ wins. Formally, $W_i(\vec s) = \{c\in M : \exists \vec s'\in S(\vec s, r_i) \text{ s.t. } f(\vec s',c)=c\}$. Also denote $W_0(\vec s) = \{c\in M : f(\vec s,c)=c\}=W_{i}(\vec s)$ for $r_i=0$, and $W_i(\vec s)=W_i(\vec s_{\vec a})$.

It is easy to see that under the multiplicative distance $c\in W_i(\vec s)$ iff $s(c) \geq (1+r_i)^{-2} s(f(\vec s))$ (see dashed line in Fig.~\ref{fig:ex}), and that similar thresholds exist for other metrics (see also Lemma~3 in \cite{MLR14_full}).  %The next key lemma, however, holds only for candidate-wise distances.%the multiplicative distance and the $\ell_\infty$ but not for other $\ell_d$ norms.  

\begin{lemma}
\label{lemma:pairs}
Every pair of possible winners are tied in some possible state. Formally, for every $b,c\in W_i(\vec s)$, there is $\vec s'\in S(\vec s,r_i)$ s.t. $b,c\in W_0(\vec s')$.
\end{lemma}
\begin{proof} %[Proof for the multiplicative distance]
Consider some $b,c\in W_i(\vec s)$.
For the multiplicative distance, let the score of all candidates except $b,c$ be $s'(a) = s(a)/(1+r_i)$, and set $s'(b) = s'(c) = \min\{s(b),s(c)\}(1+r_i)$. %\footnote{Under the $\ell_\infty$ we set $s'(b)=s'(c)=\min\{s(b),s(c)\}+r_i$, and reduce the score of all other candidates by $r_i$.}
Then $\vec s'\in S(\vec s,r_i)$, and $s'(b)=s'(c) \geq s'(a)$ for all $a\in M$.
\end{proof}

%We first show that in every strategic response, a voter always votes for her favorite possible winner. 
We denote by $a_i \step{i} a'_i$ valid local dominance steps where $a'_i\in g_i(\vec a)$ and $a'_i\neq a_i$.
 The next two lemmas characterize such LD moves.

%In the case of $\ell_\infty$, we can provide a more precise characterization.
\begin{lemma}
\label{lemma:worst_winner}
 Consider an LD move $a_i \step{i} a'_i$. Then either (a) $a_i \notin W_i(\vec s)$;  or (b) $a_i \prec_i b$ for all $b\in W_i(\vec s)$; or (c) $r_i=0$, $\{a_i,a'_i\}\subseteq W_0(\vec s)$ and $a'_i \succ_i a_i$.
\end{lemma}
\begin{proof}
%The lemma does not hold in general for arbitrary distance metrics. However it does hold for $\ell_\infty$ and for the multiplicative distance, and it is the main property we use in our proofs (so for any metric for which the lemma holds, our other results hold as well).
Suppose that $a_i , b\in W_i(\vec a)$, and $a_i \succ_i b$ (i.e., (a) and (b) are violated). 
Assume first that  $a'_i\notin W_0(\vec s)$. 
By Lemma~\ref{lemma:pairs} there is a state $\vec s'\in  S(\vec s,r_i)$ where $a_i,b$ have maximal score (possibly with other candidates), strictly above $a'_i$. W.l.o.g. $Q^+_{\vec s'}(b)<Q^+_{\vec s'}(a_i)$, as the tie-breaker does not affect the distance.
Thus $f(\vec s',a_i) = a_i, f(\vec s',a'_i) = b$. %If then $f(\vec s',c)=b$ for any $c\neq a_i$. 
Since $a_i \succ_i b$, then we have that $a_i$ $S(\vec a,r_i)$-beats $a'_i$. %, and thus $a'_i\notin D$.

The remaining case is where $a'_i\in W_0(\vec s)$ and $a_i \succ_i a'_i$. Then in the state $\vec s'$ where $a_i,a'_i$ are tied it is better to vote for $a_i$. In either case we get that $a'_i\notin D$, which is a contradiction.% every possible $\vec s'$ where  $a_i,b\in W_0(\vec s')$, we also have $a'_i\in  W_0(\vec s')$. Thus % Thus $r_i=0$ and $W_i(\vec s)=W_0(\vec s)$. \rmr{this is assuming that scores can fluctuate both up and down}
\end{proof}

\begin{lemma}
\label{lemma:best_winner}
Consider an LD move $a_i \step{i} a'_i$. Then
\begin{enumerate}
	\item $a'_i \in W_i(\vec s)$, and there is some $c\in W_i(\vec s)$ s.t. $a'_i \succ_i c$. 
	\item For a strict LD move, $a'_i = \argmin_{c\in W_i(\vec a)}Q_i(c)$. 
	\item If $a_i\notin W_i(\vec s)$, then $|g_i(\vec s)|=1$ (thus weak and strict LD coincide).
\end{enumerate}
\end{lemma}
Part~1 is the only thing we need for our later convergence results, and its proof is immediate: Consider $a^*_i = \argmin_{c\in W_i(\vec a)}Q_i(c)$. Clearly $a^*_i$ locally dominates any candidate not in $W_i(\vec a)$, thus $a'_i \in W_i(\vec s)$. If there is only one possible winner, no action dominates any other action, and there are no moves.  Part~2 is the nonatomic analog of Lemma~5.1 in \cite{MLR14}, however our proof is simpler (see Appendix~\ref{apx:distance}).

The above lemmas show that a strict LD move boils down to a simple heuristics: vote for the most preferred candidate among those whose score is above the threshold. For a weak LD move, any candidate above the threshold except the least preferred can be selected.

	%Lemma~\ref{lemma:threshold} does not hold for the EM distance and for other $\ell_d$ norms, since the score of candidates other than the winner may determine the threshold $\beta(\cdot)$. 

\newsubsec{Existence of equilibrium and convergence}
Suppose that voters start at some arbitrary state $\vec a^0$, and play repeatedly. We get a sequence of states $\vec a^0,\vec a^1,\ldots$ where in every iteration some arbitrary subset of voters may change their vote. That is, for any $t\in \mathbb N$ and any $i\in I$, either $a^{t+1}_i \in g_i(\vec a^t)$, or $a^{t+1}_i = a^t_i$. 
Our primary result states that voters always converge in the nonatomic model (see Appendix~\ref{apx:finite} for a proof and discussion of the finite case).\footnote{We assume strict preference orders for consistency with most of the social-choice literature. However we note that the theorem applies also under weak preferences. Indeed, introducing indifference in the preference relation only eliminate LD moves, and therefore never creates a cycle.}
\begin{theorem}
\label{th:LD_converge}
Any sequence of weak LD moves is finite.
\end{theorem}
\begin{proof}
We only need to show that no valid sequence may contain a cycle (see Appendix~\ref{apx:acyclic}).
Assume, toward a contradiction, that there is a cyclic path $(\vec a^t)_{t=0}^T$, and denote by $R\subseteq M$ all candidates that are part of the cycle. 
Let $s^*$ be the lowest score of any candidate in $R$ during the cycle, w.l.o.g. candidate $a^*\in R$ at time $t^*$. Thus $s^*=s^{t^*}(a^*)\leq s^t(c)$ for all $c\in R$ for every time $t\leq T$.  Consider the next step where some voters join $a^*$ (w.l.o.g. at step $t^*$), and pick an arbitrary voter $j\in I$ s.t. $a^* = a^{t^*+1}_j \neq a^{t^*}_j$. % = a_j$ (a voter that moved to $a^*$). 
Thus at step $t^*$ there is a move $a_j \step{j} a^*$ where $a_j=a^{t+1}_j\in R$. 

By Lemma~\ref{lemma:best_winner},  $a^*$ is a possible winner for voter~$j$, i.e., $a^*\in W_j(\vec s^{t^*})$. Since $s^{t^*}(c) \geq s^{t^*}(a^*)$ for all $c\in R$, we have  $R\subseteq W_j(\vec a^{t^*})$, and in particular $a_j\in W_j(\vec a^{t^*})$.

 By Lemma~\ref{lemma:worst_winner}, either $a_j$ is the least-preferred candidate for $j$ in $W_j(\vec s^{t^*})$ (Cases~I and II below), or the third category of the lemma holds. We treat the latter case separately (Case~III), so assume $a_j$ is indeed the least-preferred in $W_j(\vec s^{t^*})$. Since $R\subseteq W_j(\vec a^{t^*})$, $a_j$ is the least-preferred in $R$ as well. 

There must be some step $t^{**}$ in the cycle where  a voter of type~$v_j$ moves \emph{to} $a_j$ (w.l.o.g. voter $j$). So in $\vec s^{t^{**}}$, $a_j$ is preferred by $j$ to \emph{some} other possible winner $z$ by Lemma~\ref{lemma:best_winner}. Since $a_j$ is the least-preferred in $R$, and $a_j \succ_j z$, we have that $z \in W_j(\vec s^{t^{**}})\setminus R$.  Denote the (fixed) score of $z$ by $s(z)$. 

Case~I: $s(z) \geq s^*$. 
%We have that in particular, $s^*\leq s^{t^{**}}(a^*) < s^{t^{**}}(z) = s(z)$. 
Consider again step $t^*$. Since $s(z)\geq s^*$, we have $z\in W_j(\vec s^{t^*})$. Since $a_j$ is the least preferred possible winner in $t^*$,  we have that $z \succ_j a_j$, which is a contradiction.
\rmr{For $\ell_1$ and other metrics, it is not always true that $a_j$ is the least preferred in $W_j(\vec s^{t^*})$. It is possible, e.g. that all candidates can only be tied with the winner $f=f(\vec s^{t^*})$, and $ a^* \succ_j f \succ_j a_j \succ_j z$.}

Case~II: $s(z)<s^*$. Denote $d=a^{t^{**}}_j$, and consider the step $d \step{j} a_j$ at time $t^{**}$. Since  $d\in R$ then $s^{t^{**}}(d) \geq s^* > s(z)$, and thus $d\in W_j(s^{t^{**}})$.  By Lemma~\ref{lemma:worst_winner} (category (b) or (c)), we have that $a_j \succ_j d$. This is a contradiction since $a_j$ is the least-preferred in $R$.

Case~III: The remaining case is when there is $b\in W_j(\vec s^{t^*})$ s.t. $a_j\succ_j b$. Then by Lemma~\ref{lemma:worst_winner}, $a^*\in W_0(\vec s^{t^*})$. However since $a^*$ has minimal score, this means that $R \subseteq W_0 (\vec s^{t^*})$, i.e., all candidates in the cycle have the same score $s^*$ at  time $t^*$. Then all of $R$ must have the same score at \emph{every time} $t$, since if the score of some candidates goes up, the score of others must go down below the minimum $s^*$. This means that all of the moves in the cycle  fall under categories (b) or (c) of Lemma~\ref{lemma:worst_winner}. Thus voters only vote for more preferred candidates, which contradicts a cycle.
\end{proof}
One can argue that the uncertainty level of a voter may not remain the same throughout the game. For example, there may be less uncertainty as the game advances. %, or more uncertainty as voters learn that their prior beliefs are not so accurate. 
We note that our proof still goes through (and thus Theorem~\ref{th:LD_converge} holds) even if $W_j(\vec s^{t^*})$ and $W_j(\vec s^{t^{**}})$ are obtained via different values of $r_j$. %Thus Theorem~\ref{th:LD_converge} holds even if $r_j$ changes throughout the game. 

%Another assumption that can be relaxed is that all voters share the same prospective score vector.\footnote{Although it may be complicated to describe a formal model without this assumption.} Indeed, the beliefs of voters of other types do not play a role in our proof.

%\rmr{convergence speed?}

%The same proof, with some minor modifications, applies for atomic games where voters move one at a time. % We simply replace ``type~$j$'' with ``voter~$j$'', which is the unique voter moving in $t^*$ and $t^{**}$.

The above theorem proves convergence under very broad conditions, but does not provide much intuition as to what happens along the converging path.
 Our next result shows that when all voters have the same uncertainty level, convergence is much more structured. Our result extends a similar result in \cite{MLR14} for the finite model, but note that in our model convergence is guaranteed even when subsets of voters move simultaneously.

 Conveniently, when all voters has the same uncertainty $r$, at any state $\vec a^t$ there is just one agreed set of possible winners, denoted by $W^t = W(\vec a^t)$. We say that a move $a_i \step{i} a'_i$ is an \emph{opportunity move} if $a'_i \succ_i a_i$, and otherwise it is a \emph{compromise move}.

\begin{proposition}
\label{th:LD_truth}
Consider any non-atomic Plurality voting game with weak LD moves, where all voters have the same uncertainty level $r$.
If $\vec a^0$ is the truthful state, then for all $t$: (A) $W^{t+1} \subseteq W^{t}$; (B) the score of the winner is non-decreasing;  (C)  there are only compromise moves; and (D) any $a^t_i$ is either the most preferred candidate for $i$ in $W^t$, or it is not in $W^t$.
\end{proposition}
The proposition still holds if the uncertainty level $r$ \emph{decreases} over time.

\section{Regret Minimization}

The local dominance approach is appropriate to describe voters who are reluctant to change their vote, unless they know it cannot hurt them, a behavior that is consistent with loss aversion. A different approach to decision making under strict uncertainty is minimization of  \emph{worst case regret}, explained by \emph{risk aversion}~\cite{bell1982regret}. Intuitively, such a voter wants to avoid situations in which she could have significantly improved her utility, but failed to be pivotal.

On one hand, regret minimization is simpler than local-dominance since it does not depend on the current vote. Thus we can write $g_v(\vec s),W_v(\vec s)$ rather than $g_i(\vec s),W_i(\vec s)$.
On the other hand, regret may depend on \emph{cardinal utilities}, rather than ordinal preferences.

A cardinal utility scale is a generic function $u:M\rightarrow \mathbb R$, where $u(a)\neq u(b)$ for all $a\neq b$. A cardinal utility scale $u$ \emph{fits} order $Q\in \pi(M)$, if $u(a) > u(b)$ whenever $Q(a)<Q(b)$. We thus augment the definition of a type $v$ with a cardinal utility scale $u_v$ that fits $Q_v$.

Formally, the \emph{regret} of a type $v$ voter for voting $b$ in state $\vec s'$ is $REG_v(\vec s',b)= \max_{c\in M} u(f(\vec s',c))- u(f(\vec s',b))$. Note that $REG_v(\vec s',b)\geq 0$. 

The \emph{worst case regret} (WCR) of $i\in I$ for voting $b$ in $\vec s$ is $WCR_v(\vec s,b) = \max_{s'\in S(\vec s,r_v)}REG_v(\vec s',b)$.

\begin{definition}[Strategic move under regret minimization]
\label{def:WCR}
A WCR voter of type $v$ in profile $\vec a$ votes for a candidate $b$ minimizing $WCR_v(\vec s_{\vec a},b)$. %If $WCR_v(\vec s_{\vec a},b)=0$ for all $b\in M$,  then the voter votes arbitrarily.\footnote{We can decide, for example, that the voter votes for $q_v$, or holds her vote $a_i$.}
\end{definition}

We first characterize WCR moves. We show that (under a candidate-wise distance) we can effectively ignore the cardinal utility scale, as only the preference order affects the vote.
\begin{lemma}
\label{lemma:WCR_winner}
%Let $a_i \step{i} a'_i$ be a strict LD move in profile $\vec a$. Then $a'_i$ is the
Either $|W_v(\vec s)|=1$ (in which case all regrets are 0); or
 the \emph{unique} candidate minimizing $WCR_{v}(\vec s,c)$ is  $a^*=\argmin_{c\in W_v(\vec s)}Q_i(c)$.
\end{lemma}

From Lemmas~\ref{lemma:best_winner} and \ref{lemma:WCR_winner}, we get that regret minimization provides a partial justification for the strict variant of LD moves.
\begin{corollary}
\label{th:LD_WCR}
Let $a_i \step{i} a'_i$ be a strict LD move in state $\vec s$, then $a'_i$ is the WCR response of type $v_i$, i.e., the unique candidate $c$ minimizing $WCR_{r_i}(\vec s,c)$.
\end{corollary}
This does not mean that the WCR dynamics and the strict LD dynamics coincide:
 in an arbitrary state $\vec a$ it is possible that  $a_i$ is a possible winner that is undominated (and hence $i$ will not move under LD), yet there is another possible winner $a'
_i \succ_i a_i$ (and hence  $i$ will move to $a'_i$ under WCR).

It is an open question whether a WCR game  where all voters have the same $r$ is acyclic, yet if we add another restriction then we get that WCR moves coincide with LD moves, and in particular must converge.
\begin{proposition}
Consider any non-atomic Plurality voting game, where all voters have the same uncertainty level $r$. 
If $\vec a^0$ is the truthful state (and $|W_r(\vec a^0)|\geq 2$), %\footnote{We need assume there are at least two possible winners in $\vec a^0$, as otherwise the WCR response is not well-defined.} 
then for every time $t$ and any $i\in I$, the WCR move and the strict LD move of $i$ coincide. In particular, the WCR dynamics converges to an equilibrium.
\end{proposition}
\begin{proof}
Consider any agent $i\in I$.
By property (D) of Proposition~\ref{th:LD_truth}, we have that either $a^t_i \notin W_r(\vec s^t)$, or $a^t_i$ is the best possible winner. In the first case, by Lemma~\ref{lemma:best_winner}, there is a strict LD move, and by Corollary~\ref{th:LD_WCR}, this move coincides with the WCR move. 

In the latter case, there is no LD move for $i$ (so $a^{t+1}_i = a^t_i$ under LD), and by Lemma~\ref{lemma:WCR_winner}, $a^t_i$ minimizes worst case regret (so $a^{t+1}_i = a^t_i$ under WCR). 

Finally, since the LD dynamics converges, we get that under the conditions of the proposition (same $r$, truthful initial state), WCR converges as well.
\end{proof}
%
%
%The local dominance framework has two primary components. The first is the epistemic assumptions, namely that voters consider as ``possible'' states that are close to the prospective state. The second component is the behavioral assumption, namely that voters only leave candidates that are locally dominated. While this is a well-established approach to rationality under strict uncertainty, other prominent approaches have also been suggested. From those, perhaps the most studied one is the \emph{regret minimization} approach, according to which a decision maker will select the action that would minimize her maximal regret w.r.t. the optimal action taken in retrospect. That is, to calculate the maximal regret of action $c$ we compare the utility of voting $c$ at every possible state $\vec s'$  (denote $u(\vec s',c)$) to the best utility the voter could achieve ($\max_{b} u(\vec s',b)$). The agent then selects the action $c$ that minimizes 
%$$MaxReg_i(c,\vec s) = \max_{\vec s'\in S(\vec s,r_i)}(\max_{b\in M}u_i(f(\vec s',b))- u_i(f(\vec s',c))).$$

\newpar{Diverse population}
For the local-dominance dynamics, Theorem~\ref{th:LD_converge} shows that any game is acyclic. Since under regret minimization voters are more likely to have a strategic move, it is also more likely that cycles emerge, and even the existence of an equilibrium is not guaranteed (see Appendix~\ref{apx:regret} for proof).
%We give examples under the finite and under the nonatomic model in .
%Indeed, the our next examples shows not only that, but that a voting equilibrium may not even exist under the WCR dynamics. 

\begin{proposition}\label{th:WCR_noeq}
There is a voting game where no voting equilibrium exists under WCR dynamics.
\end{proposition}

\section{Discussion and Related Work}

We refer the reader to \cite{MLR14}, which surveyed many game-theoretic models of voting equilibrium, in particular w.r.t. various approaches to uncertainty and dominance. In addition, Meir et al. showed via extensive simulations that the equilibria reached by LD voters reproduce patterns observed in the real world, such as Duverger's Law~\cite{duverger1954political}.%and are also better for the society than the truthful outcome of Plurality.

Other concepts similar to local-dominance moves have been recently introduced in the voting literature, notably \emph{$\Pi$-manipulation}~\cite{conitzer2011dominating,reijngoud2012voter} and  \emph{'de re' manipulation} \cite{van2012strategic}. However these papers did not consider distance metrics and did not present any results on equilibrium existence or convergence.

Voting behavior based on regret minimization was considered by Ferejohn and Fiorina~\shortcite{ferejohn1974paradox}. However their model (like probability-based models) heavily relies on voters having cardinal utilities. Also, they take an extreme approach where voters do not use \emph{any} available information, and thus all states are considered possible. Another regret-based model was suggested in \cite{merrill1982strategic}, which also ignores any available information. Merril shows that under the Plurality rule uncertain voters should be \emph{truthful}, which stands in sharp contrast to behavior observed in the real world.

We see our regret minimization model as a non-probabilistic variation of the Myerson and Weber model~\shortcite{MW93}. Specifically, in the MW model the voter considers the probability of each tie to conclude her expected utility. In our WCR model the voter focuses on the \emph{most significant} possible tie, which greatly facilitates the decision making process.

Voting experiments suggest that human voting behavior is consistent with regret minimization~\cite{blais1995people,krueger2008game}, though voters may not see themselves as such. We should note that these studies have limited relevance to our work since they concern the decision \emph{whether to vote} (voter turnout), rather than the strategic decision \emph{what to vote}. Thus more experimentation is required to test the validity of such models.

A related approach is \emph{iterated regret minimization}~\cite{halpern2012iterated}, which was recently applied to voting~\cite{MW14}. This approach assumes that voters know exactly both the preferences and the decision making process of the other voters, whereas both of these assumptions are avoided in the models we studied.

A recent result on memoryless dynamics in atomic games, suggests that in any game with more than one pure equilibrium, cycles must occur if arbitrary groups of agents can move simultaneously~\cite{jaggard2011distributed}. Our result shows that this is no longer true in nonatomic games, and in fact there is a large class of games where convergence is guaranteed despite the existence of multiple equilibria.

\subsection{Conclusion and Future Directions}
We showed that in the Plurality voting system, voters who avoid locally-dominated candidates will always converge to an equilibrium, and that this result is robust to the uncertainty levels in the populations, the initial state, and the order in which voters or groups of voters play.

\newpar{Convergence rate}
In finite games without uncertainty, it is known that convergence must occur in polynomial time in the number of voters and candidates. 
The speed of convergence in non-atomic games is not easily defined. For example agents can just move in smaller and smaller masses in an infinite (acyclic) path. 
However one can ask if there is a bound on the maximal number of times a \emph{single} voter can move. For example under the conditions of Prop.~\ref{th:LD_truth}, any voter can move at most $m-1$ times, so we can say that convergence is ``fast.''
Our general convergence proof does not provide a bound of such sort, but we believe it is an interesting open question for future research. We conjecture that no agent should move more than a polynomial (in $m$) number of times until convergence occurs.

\newpar{Justifying best-response}
Consider the assumption made in \cite{MLR14}, that among all candidates dominating her current action, a voter will always select the one that is most preferred. Our paper tackles this assumption in two ways. First, we show that it is not required for convergence, and can thus be relaxed (among the other restrictions we relax). Second, we show that this assumption can be justified on the grounds that it minimizes the worst case regret of the voter.

\newpar{Experimental validation}
\rmr{add}
%Voting behavior based on local dominance relations for other voting rules have been recently formulated by several authors~\cite{conitzer2011dominating,reijngoud2012voter,van2012strategic}, however except in Plurality we are unaware of any results about equilibrium existence or properties. Further, as the action space in most voting rules is very large (permutations over candidates),
\newpar{Other voting rules}
While the definition of weak LD voters naturally extends to many other voting rules (as do the definitions in \cite{conitzer2011dominating,reijngoud2012voter,van2012strategic}),  the action space in most rules contains all permutations of $M$, and   there may be many actions that dominate a particular action. Thus the mere definitions are not very instructive as to how a voter would act. 
We hope the (local) worst-case regret minimization approach will be useful in defining reasonable voting behaviors under uncertainty for other voting rules. Finally, distance based uncertainty may prove useful in other games where there is a natural metric over action profiles. 

\section*{Acknowledgments}
The author would like to thank Omer Lev, David Parkes  and Maria Polukarov for insightful discussions and for commenting on drafts of this paper, and Harvard CRCS for the support. Some of the future directions and clarifications were pointed out by anonymous referees. 

%\bibliographystyle{plain}
%\begin{small}
%\bibliography{plurality}
%\end{small}
%\newpage
%\onecolumn
%\include{plurality.appendix}

%\end{document}

\setcounter{secnumdepth}{1}
\onecolumn
\title{Online appendices for paper \#20}
\maketitle
\onecolumn
\renewcommand{\thesection}{\Roman{section}}
\section{Characterization of LD moves}
\label{apx:proofs}
\begin{rproposition}{th:LD_truth}
Consider any non-atomic Plurality voting game with weak LD moves, where all voters have the same uncertainty level $r$.
If $\vec a^0$ is the truthful state, then for all $t$: (A) $W^{t+1} \subseteq W^{t}$; (B) the score of the winner is non-decreasing;  (C)  there are only compromise moves; and (D) any $a^t_i$ is either the most preferred candidate for $i$ in $W^t$, or it is not in $W^t$.
\end{rproposition}
\begin{proof}
The proof is very similar to the one from \cite{MLR14}.  Since the game is non-atomic the proof is somewhat simpler and also applies to group-moves.  In contrast to \cite{MLR14} (who focused on the $\ell_1$ metric), our proof also holds for \emph{weak} LD moves under candidate-wise metrics.

We show that by induction on $t$, all properties (A-D) hold. This clearly holds in $t=0$, since $a^0_i=q_i$. Now, consider some move from $\vec a^t$ to $\vec a^{t+1}$. By the induction hypothesis all properties hold until $\vec a^t$. 

Consider any move $a^t_i \step{i} a^{t+1}_i$ for some $i\in I$. 
If this is the first move of voter $i$, then clearly this is a compromise move. Otherwise, since $a^t_i$ is the most preferred candidate of $i$ in some $W^{t'}$, $t'<t$, and $a^{t+1}_i\in W^t \subseteq W^{t'}$, then $a^t_i \succ_i a^{t+1}_i$, which shows (C).
By Lemma~\ref{lemma:best_winner}, $a^{t+1}_i\in W^t$. Since $a^{t}_i \succ_i a^{t+1}_i$,  categories (b) and (c) of Lemma~\ref{lemma:worst_winner} are impossible, thus category (a) applies, and  $a^{t}_i\notin W^t$. In particular $a^t_i \neq f(\vec a^t)$, thus neither (A) nor (B) are violated at time $t$.

 Finally, by Lemma~\ref{lemma:best_winner} $a^{t+1}_i$ is the most preferred in $W^t$. Since $W^{t+1}\subseteq W^t$, (D) holds as well.
\end{proof}
The proposition still holds if the uncertainty level $r$ \emph{decreases} over time, but not if it can grow (since then it may occur that $W^{t+1} \nsubseteq W^t$).

\section{Atomic voters}
\label{apx:finite}

\paragraph{Basic notations}
We denote $[x]=\{1,2,\ldots,x\}$. 
The set of candidates is denoted by $M$, respectively, where $m=|M|$. Let $\pi(M)$ be the set of all orders over $M$.

% Thus a voter of type $v$ has preferences $Q_v\in \pi(M)$, where $Q_v(a)\in[m]$ is the rank of candidate $a\in M$, and $q_v=Q^{-1}_v(1)$ is her most-preferred candidate. 

There is a finite set of voters $I$ of size $n$.
A \emph{preference profile} $\vec Q\in\pi(M)^n$ is a collection over preferences, where $Q_i$ is the preference order of voter~$i\in I$. We denote by $v_i$ the \emph{type} of voter~$i$, which includes her preferences and possibly other information on her beliefs and behavior. Thus a population $\vec V = (v_i)_{i\in I}$ specifies the types of all voters. For a vector $\vec x = (x_j)_{j\in I}$, we denote by $\vec x_{-i}$ the partial vector that includes all entries except $x_i$. 

 We denote $a \succ_i b$ if $Q_i(a) < Q_i(b)$.

An \emph{action profile} (a.k.a. state) is a vector $\vec a=(a_i)_{i\in I}$, where $a_i\in M$ is the candidate that voter $i$ votes for. 

The Plurality voting rule $f$ allows voters to vote for a single candidate from the set $M$. Then, $f$ chooses the candidate with the highest score. 
Thus any population defines a \emph{game}, where voters' types determine their utilities. 
The outcome of the Plurality rule only depends on the total number of votes for each candidate. Formally,
we define the \emph{score vector} $\vec s_{\vec a}$ induced by action profile $\vec a$, as $s_{\vec a}(c) = \sum_{i\in I: a_i=c}w_i$, where $w_i=1$ for all $i$ (compare with the corresponding definition in the nonatomic case). 

 Every score vector $\vec s$ implicitly contains an arbitrary ``tie-breaker'' $Q^+_{\vec s}\in \QQ$.  
The \emph{winner} $f(\vec s)$ is the candidate $c\in M$ whose score $s(c)$ is maximal if such a candidate is unique. Otherwise we break ties according to $Q^+_{\vec s}$, thus $f(\vec s) = \argmin_{c\in\argmax s(c)}Q^+_{\vec s}(c)$. 
An alternative (but equivalent) way to compute the winner in the atomic case, is to think of the tie breaker as if it adds a small fraction to the score of each candidate. Formally, $f(\vec s) = \argmax_{c\in M} (s(c)+\frac{m+1-Q^+_{\vec s}(c)}{m+1})$.

\paragraph{Dynamics}
We describe the behavior of a voter $i\in I$ by a \emph{response function} $g_i:\SS \rightarrow 2^M\setminus \emptyset$. That is, a mapping from the current state to a subset of valid actions. In contrast to the atomic case, the state (score vector) considered by  voter $i$ does not contain her own action. Thus $g_i(\vec a)$ is a shorthand for $g_i(\vec s_{\vec a_{-i}})$, and a voter $i\in I$  may choose any action in $g_i(\vec a)$.
Together, $Q_i$ and $g_i$ completely define the type $v_i$ of voter $i$. 

For example, $g_i$ can be the best-response function: $g_i(\vec a)= \argmin_{c\in M}Q_i(f(\vec a_{-i},c))$. 
 \begin{definition}
\label{def:eq_NA_finite}  Let $\vec V$ be a population of voters. 
A \emph{voting equilibrium} is a state $\vec a$, where $g_i(\vec a) = \{a_i\}$ for all $i\in I$.
\end{definition}
For example, if $g_i$ is the best-response function (or better-response, which is set-valued), then a voting equilibria coincides with the Nash equilibria of the atomic game.
 
\paragraph{Local dominance}

The set $S_i(\vec s,x)$ is exactly as in the nonatomic case, i.e.
$$S(\vec s,x)=\{\vec s' : \delta(\vec s,\vec s') \leq x\},$$
however for a profile $\vec a$, $S_i(\vec a,x) = S_i(\vec s_{\vec a_{-i}},x)$. 
In particular, and in contrast to the nonatomic case, two  voters of the same type may consider different sets of possible states, if they happen to vote for different candidates in $\vec a$. For the multiplicative metric, $s'(c)$ is in the range $[\floor{ s(c) / (1+x)}, \ceil{ s(c)  (1+x)}]$.

We define the \emph{uncertainty parameter} $r_i\in \mathbb R^+$ as part of the type $v_i$. 
Given a prospective state $\vec s$,  voter $i$ believes that the actual state may be any $\vec s' \in S_i(\vec a,r_i)$. 

Since $\vec s'$ also does not contain the vote of $i$, the winner that $i$ considers in $\vec s'$ depends on her vote $a_i$. That is, $f(\vec s',a_i)$ is the winner of the profile $\vec s'$ where $a_i$ gets one additional vote. 
Note that in the nonatomic case, the vote $a_i$ is assumed to break any ties, and the tie breaker $Q^+_{\vec s'}$ is only considered if $a_i$ is not a candidate with maximal score. In contrast, in the atomic case $a_i$ simply adds one vote to $s(a_i)$, and then we apply $Q^+_{\vec s}$ to resolve any ties. Thus if several candidates $C\subseteq M$ are tied in $(\vec s,a_i)$ with maximal score, each of them wins in some possible state.

\begin{definition}[Strategic move under local dominance for finite populations]
\label{def:LD_finite}
Let $a_i$ be the vote of $i$ in profile $\vec a$.   
Let $D\subseteq M$ be the set of candidates that $S(\vec a,r_i)$-dominate $a_i$, and that are not $S(\vec a,r_i)$-dominated. If $D=\emptyset$ then $g_i(\vec a)=\{a_i\}$. Otherwise:
\begin{itemize}
	\item For a \emph{weak LD voter}, $g_i(\vec a)= D$. 
	\item For a \emph{strict LD voter}, $g_i(\vec a) = \argmin_{d\in D} Q_i(d)$.
\end{itemize}
\end{definition}
In other words,  a strict LD voter votes  the most preferred  candidate that locally-dominates her current choice (if such exists). The definition of a weak LD voter is much more permissive, and does not restrict the voter to select the most preferred candidate in $D$.

Note that the definition looks identical to Def.~\ref{def:LD} from the nonatomic model. The only difference is that $S_i(\vec a,r_i)$ in the finite case depends on the identity (on the current vote) of $i$, and not only on its type. Note however that $g_i$ depends on $a_i$ both in the finite and nonatomic cases.

Definition~\ref{def:LD_finite} for strict local dominance almost coincides with the definition in of local dominance in \cite{MLR14}, but for one difference:  We require $D$ to only contain undominated candidates whereas no such restriction 	appeared in \cite{MLR14}. Note that without this restriction, a single voter can move, and then immediately move again, which does not make much sense. 
%Note that $g_i$ does not use the entire action profile $\vec a$, but only $a_i$ and $\vec s$.

\paragraph{Characterization of LD moves}

We say that candidate $c$ is a \emph{possible winner for $i$ in state $\vec a$} if there is a possible state where $c$ wins. Formally, $W_i(\vec a) = \{c\in M : \exists \vec s'\in S_i(\vec a, r_i) \text{ s.t. } f(\vec s',c)=c\}$.

Note that in contrast to the nonatomic case, different voters of the same type may consider different sets of possible winners. This is the main advantage of the nonatomic model.

It is easy to see that under the multiplicative distance $c\in W_i(\vec a)$ iff $s_{\vec a_{-i}}(c) \geq (1+r_i)^{-2}  s_{\vec a_{-i}}(f(\vec s))$. 
The full proof of Lemma~\ref{lemma:pairs} (in Appendix~\ref{apx:distance}) that also covers the finite voters case, for any candidate-wise metric.

The proofs of Lemmas~\ref{lemma:worst_winner} and \ref{lemma:best_winner} are the same for the finite case, one we replace the notation $S(\vec a,r_i)$ with $S_i(\vec a,r_i)$.

\paragraph{Convergence}
%
%\begin{lemma}
%\label{lemma:pairs_finite}
%Every pair of possible winners are tied in some possible state. Formally, for every $b,c\in W_i(\vec a)$, there is $\vec s'\in S_i(\vec s,r_i)$ s.t. $b,c\in W_0(\vec s')$.
%\end{lemma}
%\begin{proof}
%Consider some $b,c\in W_i(\vec a)$. Thus $s^-(b),s^-(c) \geq (1+r_i)^{-2} s^-(a)$ for all $a\in M$.
%Let the score of all candidates except $b,c$ be $s'(a) = \floor{s^-(a)/(1+r_i)}$, and set $s'(b) = s'(c) = \ceil{\min\{s^-(b),s^-(c)\}(1+r_i)}$. Then $\vec s'\in S_i(\vec a,r_i)$, and 
%$$s'(b)=s'(c) \geq (1+r_i)s^-(b) \geq s^-(a) / (1+r_i) \geq s'(a),$$
 %for all $a\in M$.
%\end{proof}
Recall that when several voters (in the finite case) may move simultaneously, cycles may emerge even without uncertainty at all~\cite{MPRJ:2010:AAAI}. A simple example is when $s(a)=1, s(b)=s(c)=0$, and we add two voters that favor $b,c$ to $a$. If they start by one voting to $b$ and the other to $c$, then each time they both play they will just switch locations. 

The proof of Theorem~\ref{th:LD_converge} fails in the finite case where it uses the fact that $s^{t^*}(a') \geq s^{t^*}(a^*) = s^*$ and thus the deserted candidate $a_j$ is a possible winner for $j$. In the finite case this does not hold, since the high score of $a_j$ depends on the vote of $j$, and $j$ herself does not consider $a_j$ as a possible winner. 

We next show that the theorem still holds once restricted to singleton moves. That is, we consider all valid sequences $\vec a^0,\vec a^1,\ldots$ where $a^{t+1}_i \in  g_i(\vec a^t)$ for some voter $i\in I$, and $a^{t+1}_i = a^t_i$ for all other voters.

As we can see, in the finite case we need to cover some more issues so the proof is somewhat longer, despite the restriction to singleton moves.

\begin{theorem}
\label{th:LD_converge_finite}
Any sequence of (singleton) weak LD moves is finite.
\end{theorem}
\begin{proof}
Clearly in the finite case every sequence is either finite or contains a cycle.
Assume, toward a contradiction, that there is a cyclic path $(\vec a^t)_{t=0}^T$. Let $R\subseteq M$ be all candidates that are part of the cycle, and let $I(R)\subseteq I$ be the voters that participate in the cycle.

Let $s^*$ be the lowest score of any candidate in $R$ during the cycle. That is, we find $a^*\in R$ and $t^*$ s.t. $s^*=s^{t^*}(a^*)$ is minimal. 

By definition, $s^*=s^{t^*}(a^*)\leq s^t(c)$ for all $c\in R$ for every time $t\leq T$.  Consider the next step where some voter $j$ joins $a^*$ (w.l.o.g. at step $t^*$), and pick an arbitrary voter $j\in I$ s.t. $a^* = a^{t^*+1}_j \neq a^{t^*}_j = a_j$ (a voter that moved to $a^*$). Thus at step $t^*$ there is a move $a_j \step{j} a^*$ for some $a_j\in R$. 

By Lemma~\ref{lemma:best_winner},  $a^*$ is a possible winner for voter~$j$, i.e., $a^*\in W_j(\vec s^{t^*})$. 
We facilitate the notation by writing $\vec s_{-j}=\vec s_{\vec a_{-j}}$, which are the scores from the point of view of $j$ (i.e., omitting the vote of $j$).

Since $s^{t^*}(a') \geq s^*$ for all $a'\in R$, in particular $s^{t^*}_{-j}(a_j)+1  \geq s^*$. Suppose first that $s^{t^*}_{-j}(a_j)< s^*$. Then after the  move $s^{t^*+1}(a_j) = s^{t^*}_{-j}(a_j)< s^*$, in contradiction to the selection of $s^*$ (note that this only applies to singleton moves, as no other voter can move to $a_j$ at time $t^*$). Thus $s^{t^*}_{-j}(a_j) \geq s^* = s_{-j}^{t^*}(a^*)$, which means that $a_j\in W_j(\vec a^{t^*})$. Note that for all other candidates $c\in R\setminus \{a_j\}$, $s^{t^*}_{-j}(c) = s^{t^*}(c) \geq s^*$ as well, thus $R\subseteq W_j(\vec s^{t^*})$. The remainder of the proof is almost identical to the nonatomic case, but we write it for completeness.

 By Lemma~\ref{lemma:worst_winner}, either $a_j$ is the least-preferred candidate for $j$ in $W_j(\vec s^{t^*})$ (Cases~I and II), or the third category of the lemma holds. We treat the latter case separately (Case~III), so assume $a_j$ is indeed the least-preferred in $W_j(\vec s^{t^*})$. Since $R\subseteq W_j(\vec a^{t^*})$, $a_j$ is the least-preferred in $R$ as well. 

There must be some step $t^{**}$ in the cycle where  a voter of type~$v_j$ moves \emph{to} $a_j$ (w.l.o.g. voter $j$). So in $\vec s^{t^{**}}$, $a_j$ is preferred by $j$ to \emph{some} other possible winner $z$ by Lemma~\ref{lemma:best_winner}. Since $a_j$ is the least-preferred in $R$, and $a_j \succ_j z$, $z \notin R$. Thus there is some $z\in W_j(\vec s^{t^{**}})\setminus R$ such that $z\prec_j a_j$. Denote the (fixed) score of $z$ by $s(z)$. 

Case~I: $s(z) \geq s^*$. 
%We have that in particular, $s^*\leq s^{t^{**}}(a^*) < s^{t^{**}}(z) = s(z)$. 
Consider again step $t^*$. Since $s(z)\geq s^* = s_{-j}(a^*)$ and $a^*\in W_j(\vec a^{t^*})$, we have $z\in W_j(\vec a^{t^*})$. Since $a_j$ is the least preferred possible winner in $t^*$,  we have that $z \succ_j a_j$, which is a contradiction.
\rmr{For $\ell_1$ and other metrics, it is not always true that $a_j$ is the least preferred in $W_j(\vec s^{t^*})$. It is possible, e.g. that all candidates can only be tied with the winner $f=f(\vec s^{t^*})$, and $ a^* \succ_j f \succ_j a_j \succ_j z$.}

Case~II: $s(z)<s^*$. Denote $d=a^{t^{**}}_j$, and consider the step $d \step{j} a_j$ at time $t^{**}$. Since  $d\in R$ then $s^{t^{**}}_{-j}(d) \geq s^* > s(z)$, and thus $d\in W_j(a^{t^{**}})$ (as $z\in W_j(a^{t^{**}})$).  By Lemma~\ref{lemma:worst_winner} (category (b) or (c)), we have that $a_j \succ_j d$. This is a contradiction since $a_j$ is the least-preferred in $R$.

Case~III: The remaining case is when there is $b\in W_j(\vec a^{t^*})$ s.t. $a_j\succ_j b$. Then by category (c) of Lemma~\ref{lemma:worst_winner}, $a^*\in W_0(\vec a^{t^*})$. However since $a^*$ has minimal score, this means that $R \subseteq W_0 (\vec a^{t^*})$, i.e., all candidates $c\in R$ have the same score $s_{-j}^{t^*}(c)=s^*$ at  time $t^*$, and there is exactly one candidate ($a_j$) that has the additional vote of voter~$j$. Since throughout the cycle the score of candidates in $R$ may not drop below $s^*$, we have that all moves in the cycle are from the single candidate $c^t\in R$ with score $s^t(c^t)=s^*+1$ to a candidate with score $s^*$. This means that the deserted candidate $c^t$ is always a possible winner for the mover. Consequently, all of the moves in the cycle  fall under categories (b) or (c) of Lemma~\ref{lemma:worst_winner}. Thus voters only vote for more preferred candidates, which contradicts a cycle.
\end{proof}

\paragraph{Best response and better response}
In \cite{MPRJ:2010:AAAI} a distinction was made between ``better-response'' (any move that improves the winner), ``best-response'' (any move that leads to the selection of the best possible winner); and ``unique best-response'' (or restricted best-response), which requires in addition that the voter votes for the selected candidate. 
It was shown that only unique best-response guarantees convergence, whereas cycles may occur under better-response. It may thus seem that our result contradicts the one in \cite{MPRJ:2010:AAAI}. 

However, a similar distinction can be made between ``better-response'' and ``restricted better-response''. In \cite{MLR14_full} it was shown that local-dominance  naturally restricts the response in a similar way---a candidate that is not a possible winner is always dominated. In the special case of $r=0$ the set $D$ is exactly the set of restricted better-responses. Thus it is not the restriction to best response that leads to convergence, but rather the restriction to vote for the candidate the voter is trying to promote.

The connection between local-dominance and restricted better-response for $r=0$ only holds in our neutral tie-breaking, but it is easy to verify that restricted better-response converges also in the lexicographic tie-breaking model of \cite{MPRJ:2010:AAAI}.

\section{Convergence and Acyclicity}
\label{apx:acyclic}
Without restricting the resolution $\eps$, there may be infinite \emph{acyclic} paths, even with two candidates and one type of voters. For example consider a population where all voters prefer $b$ over $a$, but initially all vote to $a$. That is, there is just one type of voters and $s^0(a)=1,s^0(b)=0$. Suppose that in step $t$, half of the remaining $a$ voters are chosen to play, and move to $b$. This entails an infinite acyclic path.

To avoid such infinite paths, and also to allow for more intuitive notation, we assume that there is some bounded granularity level $\eps$, and that sets of voters of the same type with mass $\eps$ always move together.\footnote{Note that this is different than a coalitional manipulation, where voters are coordinated. It is possible that all voters will be less satisfied after the move.} 

Thus a state $\vec a$ is $\eps$-valid if $a(v,c)$ is an integer multiple of $\eps$, and a move from state $\vec a^t$ to state $\vec a^{t+1}$ is $\eps$-valid, if (I): both states are $\eps$-valid; (II) $\vec a^{t+1}$ is attained from $\vec a^t$ by taking each set $i\in I$, and set either $a^{t+1}_i = a_i$, or $a^{t+1}_i \in g_i(\vec a)$. 

Finally, for any $\eps>0$, there is only a finite number of $\eps$-valid states. Thus there cannot be cycles of infinite length.
\begin{observation}
If a voting game has no finite cycles, then every $\eps$-valid path, for any $\eps>0$, is finite.
\end{observation}

Finally, the assumption that all voter sets have the same size $\eps$ simplifies the notation but is not necessary. 

Clearly valid moves can be defined without referring to ``identified'' voters, although the definition is quite tedious. 
Consider a valid finite cycle of states (not necessarily $\eps$-valid for any $\eps$). Since there is a finite number of types and a finite number of states in the cycle, we can still partition the voters $[0,1]$ to subsets $I$, s.t. all voters in set $i\in I$ have the same type $v_i$ and the same sequence of actions $a^1_i,\ldots,a_i^T$. Thus we get identified voters in retrospect. The mass of each such set $i\in I$ my not be rational so there may not be common unit size $\eps$, but our convergence proof works just the same. 
%
%
%We say that profiles $\vec a,\vec a'$ are \emph{$\eps$-distinct}, if $|\vec a - \vec a'|=\eps \cdot d$ for some $d\in \mathbb Z_+$.  A \emph{$\eps$-path} is a sequence of $\eps$-distinct profiles $(\vec a^t)_{t=0}^T$, s.t. for every $t<T$ and every $i\in I(\vec a^t)$, either $a^{t+1}_i = a^t_i$, or $a^{t+1}_i \in g_i(\vec a^t)$. In words, any subset of voters (with a positive mass that is a multiple of $\eps$) may change their vote between consequent states, as long as they all make valid moves according to their response function.
%
%A game is a \emph{acyclic}, if there are no finite cyclic paths, i.e. paths $(\vec a^t)_{t=0}^T$ where $\vec a^T=\vec a^0$, $\vec a^{t+1} \neq \vec a^{t}$ for all $t<T$, and $a_i^{t+1} \in g_i(\vec a^t)$ for all $t<T,i\in I$. 
%\begin{lemma}
%Let $G$ be an acyclic game, then for \emph{any} $\eps>0$, every path of valid moves is finite.
%\end{lemma}
%\begin{proof}
%Let $\eps>0$ and $\vec a^0$ be an initial profile. Note that there is only a finite number of $\eps$-distinct states that include $\vec a^0$. Thus every $\eps$-path starting from $\vec a^0$ is either finite or cyclic. By the premise there are no cyclic paths at all, and in particular no cyclic $\eps$-paths. Thus all $\eps$-paths are finite.
%\end{proof}
 %Thus in an acyclic game, every sufficiently long path reaches an equilibrium, regardless of the initial profile, who or how many voters play, or what valid responses are taken.
%
%

\section{Voter influence}
\label{apx:influence}
In the atomic case the meaning of $f(\vec s,c)$ is clear---it is the outcome of adding one more vote to candidate $c$ in state $\vec s$. However changing the vote of a ``single'' non-atomic voter does not change the score. We remedy this by assuming each voter considers her own weight as \emph{tends to zero} but not quite zero. Formally, denote by $\vec s^{\rightarrow c}_\eps$ the state that is attained from $\vec s$ if we add to $\vec s$ a mass of $\eps>0$ voters voting for $c$. That is, $s^{\rightarrow c}_\eps(c) = s(c)+\eps$, and $s^{\rightarrow c}_\eps(b)=s(b)$ for any other candidate.

Finally, we define (in the non-atomic case) the modified outcome $f(\vec s,c)$ as $\lim_{\eps\rightarrow 0} f( \vec s^{\rightarrow c}_\eps)$. To see that the limit is well defined, note that it is sufficient to consider any $0<\eps < \min_{b,c\in M}|s(b)-s(c)|$.

By continuity of the distance metric, if there are states where two different candidates win, then there is also a state where they are tied. Thus voter $i$ believes she can affect the outcome if only if she can act as a tie-breaker between candidates in some possible state. More formally, suppose that voter $i$ votes $i$ in state $\vec s$, and consider $S=S(\vec s,r_i)$.
\begin{lemma}
\begin{enumerate}
  \item if $f(\vec s') = c$ for all $s'\in S$, then $f(\vec s',a_i)=c$ (that is, $i$ cannot change the outcome).
	\item if several candidates $C \subseteq M$ are tied (with maximal score) in some $\vec s'$, and $a_i\in C$, then $f(\vec s',a_i)=a_i$.
	\item if several candidates $C \subseteq M$ are tied (with maximal score) in some $\vec s'$, and $a_i\notin C$, then any candidate in $C$ wins in some state $\vec s''\in S$. 
\end{enumerate}
\end{lemma}
\begin{proof}
1. For all $\vec s'\in S$,   $s'(c) > s'(a_i)+ \min_{b,c\in M}|s(b)-s(c)|> s'(a_i)+\eps$ for some $\eps>0$. Thus $f(\vec s',a_i)=\lim_{\eps\rightarrow 0} f( \vec s'^{\rightarrow a_i}_\eps) = c$.

2. For all $c\in M\setminus\{a_i\}$, $s'(a_i)\geq s'(c)$ and thus $s'^{\rightarrow a_i}_\eps(a_i) >  s'(a_i) \geq s'(c) = s'^{\rightarrow a_i}_\eps(c)$. 

3. For $c\in C$, consider the state $\vec s''$ where $Q^+_{\vec s''}(c)=1$. Then the score of $a_i$ in $\vec s''^{\rightarrow a_i}_\eps$ is strictly lower than the score of $c$, as in 1. Thus $C$ have maximal score in $\vec s''^{\rightarrow a_i}_\eps$, and $c$ wins due to the tie-breaking advantage.
\end{proof}

\section{Other Distance Functions}
\label{apx:distance}
\subsection{Candidate-wise distances}

\begin{rlemma}{lemma:pairs}
Every pair of possible winners are tied for victory in some possible state. Formally, for every $b,c\in W_i(\vec s)$, there is $\vec s'\in S(\vec s,r_i)$ s.t. $b,c\in W_0(\vec s')$.
\end{rlemma}
\begin{proof} [Proof for any candidate-wise distance]
In this proof we denote the score of $c$ by $s(c)$.
Recall that $\delta(\vec s,\vec s') = \max_c \delta'(s(c),s'(c))$. 

Consider some $b,c\in W_i(\vec s)$.
Let the score of all candidates except $b,c$ be $s'(a) = \min\{z : \forall a\notin\{b,c\}, \delta'(s(a),z) \leq r_i\}$.  
 Let $h(b) = \max\{z: \delta'(s(b),z) \leq r_i\}$ and $h(c) = \max\{z: \delta'(s(c),z) \leq r_i\}$, then set $s'(b)=s'(c)=\min\{h(b),h(c)\}$.
Then $\vec s'\in S(\vec s,r_i)$ by definition. Also, we argue that $b,c$ are tied with maximal score in $\vec s'$. Indeed, since $b\in W_i(\vec s)$, we know that there is some state $\vec s_b\in S(\vec s,r_i)$ s.t. $f(\vec s_b)=b$. We know that $s_b(b)\leq h(b)$, and $s_b(a)\geq s'(a)$ for all other candidates. Since the same holds for $c$ w.r.t. some state $\vec s_c$, we get that in $\vec s'$ either $b$ or $c$ have maximal score. However by construction $s'(b)=s'(c)$ so they are tied.

Note that the proof also covers the finite case.
\end{proof}
 As this is our only proof that makes use of the properties of the distance function, all of our results hold for any candidate-wise distance.

\begin{rlemma}{lemma:best_winner}
Consider an LD move $a_i \step{i} a'_i$. Then
\begin{enumerate}
	\item $a'_i \in W_i(\vec s)$, and there is some $c\in W_i(\vec s)$ s.t. $a'_i \succ_i c$. 
	\item For a strict LD move, $a'_i = \argmin_{c\in W_i(\vec a)}Q_i(c)$. 
	\item If $a_i\notin W_i(\vec s)$, then $|g_i(\vec s)|=1$ (thus weak and strict LD coincide).
\end{enumerate}
\end{rlemma}

\begin{proof}
 Consider $a^*_i = \argmin_{c\in W_i(\vec a)}Q_i(c)$. Clearly $a^*_i$ dominates any candidate not in $W_i(\vec a)$, thus $a'_i \in W_i(\vec s)$. 

If there is only one possible winner, no action dominates any other action, and there are no moves. The fact that $a'_i\neq a_i$ entails that there is some possible state where $a'_i$ is tied with some other possible winner $c$. If $a'_i$ is the least preferred possible winner, then it is dominated by $a^*_i$, which again means there would be no move. This completes part~1.

Next for part~2. Consider the set of candidates $D$ in Def.~\ref{def:LD}. Since $a'_i\in D$, it is non-empty. We get from part~1 that $D \subseteq W_i(\vec a)$. It is left to show that  $a^*_i$ is in $D$, as this implies $a^*_i=a'_i$. 

Indeed, assume that $a'_i\neq a^*_i$. By Lemma~\ref{lemma:pairs} and part~1,  there is a state $\vec s' \in S(\vec a,r_i)$ where $ a^*_i$ is tied for victory with some other possible winner $c$, and $Q^+_{\vec s'}(c)=1$. Thus $f(\vec s', a^*_i)= a^*_i \succ_i c = f(\vec s',a_i)$, which means $a^*_i$ $S(\vec s,r_i)$-beats $a_i$.

We consider the three categories of Lemma~\ref{lemma:worst_winner}. (a) If $a_i\notin W_i(\vec s)$, then in every $\vec s'\in S(\vec s,r_i)$ voting for $a^*_i$ can only improve the outcome from $f(\vec s')\in W_i(\vec a)$ to $a^*_i$;
(b) If $a_i$ is the worst possible winner, then  $a_i$ does not beat any other candidate;
(c) There is no uncertainty except on the tie-breaker (i.e. $s'(c)=s(c)$ for all $c$, but $Q^+_{\vec s'}$ may differ), and $a^*_i,a_i\in W_0(\vec s)$. Thus in any possible state $f(\vec s',a^*_i)=a^*_i \succ_i a_i = f(\vec s',a_i)$. In all cases $a_i$ does not $S(\vec s,r_i)$-beat $a^*_i$, and thus $a^*_i\in D$.

For the third part, if $|W_i(\vec s)|=1$, then $i$ is non-pivotal and no action dominates any other action. Thus $g_i(\vec s) = \{a_i\}$ (no move). Otherwise, consider some $b\in W_i(\vec a)$ that is not $a^*_i$. Then by Lemma~\ref{lemma:pairs} there is a possible state $\vec s'$ where $b,a^*\in W_0(\vec s')$ with $Q^+_{\vec s'}(a^*_i)=1$. Then $f(\vec s',b)=b \prec_i a^*_i = f(\vec s',a_i)$, so $a_i$ beats $b$, which means $b$ cannot dominate $a_i$, and thus $b\notin g_i(\vec s)$.  
\end{proof}

\subsection{Non-candidate-wise distance functions}
For general distance-based relations, the set of possible winners alone is insufficient to determine which pairs of candidates can be tied. That is, Lemma~\ref{lemma:pairs} is violated, and thus our other proofs do not hold as well.

Under the $\ell_1$ distance, $\delta(\vec s,\vec s') = \sum_{c\in M}|s(c)-s'(c)|$. 
 Consider for example the $\ell_1$ metric, $r_i=5$, and four candidates $\{a,b,c,d\}$ whose scores are $\vec s_1 = (10,9,7,6)$. While all candidates are possible winners, the last two cannot be tied for victory. However if scores are $\vec s_2 = (10,6,6,6)$ then every pair can be tied. 

It may still hold that voting games converge for other distance functions, and we leave this as an open question for future research.

The example above can also demonstrate that cardinal utilities matter under the WCR dynamics. Suppose that the utilities for the four candidates are $\vec u_i = (3,4,5,0)$ (i.e. the third candidate is best, the fourth is worst). 
We can compute the WCR of every candidate in  by looking on all pairs that can be tied for victory. In $\vec s_2$ we have $WCR_i(\vec s,c) = \max\{4-0,3-0\} = 4$, whereas the WCR of any other action is $5$. Thus $c$ minimizes WCR. However in $\vec s_1$, the tie between $c,d$ is impossible. Thus $WCR(\vec s_1,b)=\max\{3-0,5-3\}=3$, whereas $WCR(\vec s_1,c)= WCR(\vec s_2,c) =  4$. Note that in this case the voter will not vote for her most favorite candidate ($c$), even though it is a possible winner. Finally, note that if we change $u(d)$ to 2.5 (i.e., without changing ordinal preferences $Q_i$), then $c$ is once again selected in $\vec s_1$ under WCR. 

Thus whether voters are sensitive to cardinal utilities, strongly depends on the uncertainty model we are assuming, and in particular on the distance function.

\section{Regret Minimization}
\label{apx:regret}
\begin{rlemma}{lemma:WCR_winner}
%Let $a_i \step{i} a'_i$ be a strict LD move in profile $\vec a$. Then $a'_i$ is the
Either $|W_v(\vec s)|=1$ (in which case all regrets are 0); or
 the \emph{unique} candidate minimizing $WCR_{v}(\vec s,c)$ is  $a^*=\argmin_{c\in W_v(\vec s)}Q_i(c)$.
\end{rlemma}
\begin{proof}
If there is a unique possible winner $d$ then the $f(\vec s',b)=d$ for all $\vec s'\in S(\vec s,r_v), b\in M$, and thus $WCR_v(\vec s,b) = 0$. Thus suppose that $a^*,b \in W_v(\vec s)$, where $b$ is the least-preferred for type $v$.

 We classify possible states into two classes $S_1,S_2$. If $f(\vec s',a^*) = a^*$ we say that $\vec s'\in S_1$, and otherwise $\vec s'\in S_2$.

In $\vec s'\in S_1$, $REG_v(\vec s',a^*) = 0 \leq REG_v(\vec s',c)$ for all $c\in M$. 
Recall that by Lemma~\ref{lemma:pairs}, any two possible winners can be tied for victory. 
Thus in some $\vec s'\in S_1$, $a^*,b\in W_0(\vec s')$ and $Q^+_{\vec s'}(b)=1$. Then for any $c\neq a^*$, $REG_v(\vec s',c) \geq u(f(\vec s',a^*)) - u(f(\vec s',c)) = u(a^*) - u(b)$.

In $\vec s'\in S_2$, $f(\vec s',a^*) = d \neq a^*$, then $d\in W_0(\vec s')$, $Q^+_{\vec s'}(d)=1$, and $REG_v(\vec s,a^*) = \max_{c'\in W_0(\vec s')}u(c')-u(d)$, since a voter can only bring about the election of a candidate that is tied for victory. Thus $REG_v(\vec s',a^*) \leq |u(c')-u(b)|$ for some $c',b\in W_0(\vec s'), c' \neq a^*$.

 Thus $WCR_v(\vec s,a^*) \leq |u(c')-u(b)|$ for some $c'\in W_v(\vec s), c \neq a^*$, which means that for all $c\in M$, $WCR_v(\vec s,a^*) < u(a^*) - u(b) \leq WCR_v(\vec s,c)$.
\end{proof}

In \cite{MW93}, it is shown that a voter can calculate the expected utility of voting for $c$, by averaging $u_i(c)-u_i(b)$ over all candidates $b$, weighted by the probability that $b,c$ are tied. Lemma~\ref{lemma:WCR_winner} demonstrates that under the regret minimization scheme, voters apply similar logic (only consider possible ties). However instead of taking a weighted average, the voter only considers the maximum.

\begin{rproposition}{th:WCR_noeq}
There is an example of a non-atomic voting game, where no voting equilibrium exists under WCR dynamics.
\end{rproposition}
\begin{proof}
There are $6$ candidates, $\{a,b,c_1,c_2,c_3,d\}$. 
We set the base scores of candidates to $\ol{\vec s}=(12,6,0,0,0,12)$. That is, scores due to non-strategic voters. We use the $\ell_\infty$ norm.

We introduce four types of strategic voters: $v$ and $(w_j)_{j=1}^3$. 
The preferences are $Q_v: b \succ a \succ c_i \succ d$ for all $i$, and $Q_{w_j}:c_j \succ b \succ a \succ d \succ c_i$ for $i\neq j$. We also set $r_v = 2,r_w = 6$. There are $1.5$ units of type~$v$ voters, and $1.2$ units of each type~$w_j$ voters. We note that under the $\ell_\infty$ metric, $W_r(\vec s)$ contains all candidates whose score is at least $s(f(\vec s))-2r$. Also note that the total mass of all voters is $5.1$.

%We have a cycle $(b,c) \step{i} (a,c) \step{j} (a,b) \step{i}
%(b,b) \step{j} (b,c)$. 
%This is since: (1) $b\notin W_1$ but $i$ is pivotal between $a,d$; (2) $c \notin W_4$ since $s(a)=6$, but $j$ is pivotal between $b,a$; (3) $b\in W_1$ (although does not dominate $a$); (4) $c\in W_4$ since $s(a)=5$. 
We show that in any profile $\vec a$ there is some WCR move. We denote by $x_e$, $e\in M$ the fraction of voters (of any type) voting for candidate $e$ in $\vec a$.  We observe that $s(a)\leq 12+5.1 < 18 \leq s(b) + 2r_w$, thus $b$ is always a possible winner for type~$w_j$ voters.
\begin{enumerate}
	\item If there are type~$w_j$ voters on $d,a$ or $c_i$ ($i\neq j$), they prefer $b$, since $b\in W_w(\vec a)$ and $b\succ_w d,a,c_i$. 
	%\item If there are type~$w$ voters on $a$ (and none on $d$), they prefer $b$.
	\item Similarly, if there are type~$v$ voters on $d$ or any $c_i$, they prefer $a$ (w.l.o.g. there are no type~$w$ agents on $d$, so $a\in W_v(\vec a)$).
	\item A type~$v$ voter votes for $b$ if $b\in W_v(\vec a)$ and otherwise for $a$.
	\item A type~$w_j$ voter votes for $c_j$ if $c_j\in W_w(\vec a)$ and otherwise for $b$.
	%If $x_a>x_{c_j}$, then $s(a)>s(c_j)+2r_w$. Then a type~$w_j$ voter choose $b$ (since $c_j\notin W_w(\vec a)$) and otherwise prefers $c_j$.
	%\item If $x_b<x_a+2$ then $s(b)=6 +x_b <  8+x_a = s(a)-2r_v$. Thus $b\notin W_v(\vec a)$ and  a type~$v$ choose $a$. Otherwise choose $b$.
\end{enumerate}
Assume toward a contradiction that $\vec a$ is an equilibrium under WCR. We know by 1 and 2 that $x_d=0$, $x_{c_j}\leq 1.2$ for all $j$.
We observe that $b\in W_v(\vec a)$ iff the inequality in $6 + x_b +r_v = s(b) +r_v \geq s(a)-r_v = 12+x_a - r_v$ holds. That is, iff $x_b \geq  x_a +12-6-4 = x_a+2$. Similarly, $c_j\in W_w(\vec a)$ iff $s(c_j) + r_w \geq s(a)-r_w$, which means iff $x_{c_j} \geq x_a$.

Suppose first that $b\in W_v(\vec a)$. Then all type~$v$ voters are on $b$, which means $x_a=0,x_b\geq 1.5$. This entails that $c_j\in W_w(\vec a)$ for all $j$, since $x_{c_j}\geq 0 =x_a$. We conclude that all $w_j$ voters are on $c_j$, which means $x_b \leq 1.5$. However we get that 
$$s(b) = \ol{s}(b)+x_b \leq 6+1.5 = 7.5 < 8 = s(a)-2r_v,$$
so we get that $b \notin W_v(\vec a)$. A contradiction. 

Thus suppose that $b \notin W_v(\vec a)$. Then all type~$v$ voters are on $a$, which means $x_a=1.5 > x_{c_j}$ for all $j$. This entails that $c_j\notin W_w(\vec a)$ for all $j$, since $x_a = 1.5 > 1.2 \geq x_{c_j}$. We conclude that all $w_j$ voters are on $b$, which means $x_b \geq 3.6$. However we get that 
$$s(b) = \ol{s}(b)+x_b \geq 6+3.6 = 9.6 > 9.5 = s(a)-2r_v,$$
so we get that $b \in W_v(\vec a)$. A contradiction again. 
\end{proof}

\begin{proposition}
There is an example of a atomic population $\vec V$ with four candidates, where no voting equilibrium exists under WCR dynamics.
\end{proposition}
\begin{proof}
We set the base scores of candidates to $\ol{\vec s}=(9,4,0,10)$. That is, scores due to non-strategic voters. We use the $\ell_\infty$ norm.

We introduce two strategic voters $i$ and $j$. 
The preferences are $Q_i: b \succ a \succ c \succ d$, $Q_j:c \succ b \succ a \succ d$, with lexicographic tie-breaking over candidates.
We set the uncertainty levels to $r_i = 1$, $r_j=4$. Thus a candidate is in $W_i$ if it needs more 3 votes (including $i$) to win, and in $W_j$ if it needs more $9$ votes.

 We first show that there is a cycle 
$$\vec a^1 = (b,c) \step{i} \vec a^2 = (a,c) \step{j} \vec a^3 = (a,b) \step{i} \vec a^4 =(b,b) \step{j} (b,c) = \vec a^1.$$

We can write the same cycle in terms of score vectors:
$$\vec s^1 = (\mathbf{9},5,1,\mathbf{9}) \step{i} \vec s^2 = (\mathbf{10},\mathbf{4},1,\mathbf{9}) \step{j} \vec s^3 = (\mathbf{10},5,0,\mathbf{9}) \step{i} \vec s^4 =(\mathbf{9},\mathbf{6},\mathbf{0},\mathbf{9}) \step{j} (9,5,1,9) = \vec s^1.$$

We justify the moves as follows by Lemma~\ref{lemma:WCR_winner} and the following observations for each move: (1) $b\notin W_i(\vec s^1)$ but $i$ is pivotal between $a,d$; (2) $c \notin W_j(\vec s^2)$ since $s(a)=6$, but $j$ is pivotal between $b,a$; (3) $b\in W_i(\vec s^3)$; (4) $c\in W_j(\vec s^4)$ since $s(a)=5$. 

Note that the third move by agent~$i$ cannot be an LD move, since $b$ does not dominate $a$.

Also, all other profiles are never stable: $d$ is the last candidate of both voters and thus has maximal regret in every state, so no voter votes for $d$. Voter $i$ never votes for $c$ since it is never a possible winner. Voter~$j$ never votes for $a$ since $b$ is always a possible winner and $b \succ_j a$.
\end{proof}

\section{Problematic Proof in \cite{MLR14}}
\label{apx:flaw}
Proposition~5.5 in \cite{MLR14} states that from any initial state in voting games with a finite population, there is some path that converges to an equilibrium. More specifically, that if opportunity (moves to a more preferred candidate) moves always precede  compromise moves then convergence is guaranteed. The Proposition extends this claim to subsets of agents that move simultaneously, provided that eventually enough singleton steps occur.
The proof is omitted from \cite{MLR14}, but appears in the full version \cite{MLR14_full}.

Ignoring the complications of simultaneous moves, the proof shows that after the first time there are no opportunity moves, then the score of the winner must monotonically increase until convergence. This relies on the observation from a previous proof (about convergence from the truthful state) that only voters who vote for candidates who cannot win may move. However this observation is no longer true with arbitrary initial state. 

Consider the following example. Four candidates $\{a,b,c,d\}$ with fixed scores $(5,3,4,1)$. We add two voters with $r=2$ (we use $\ell_1$ metric, but similar examples can be constructed for any metric). $i$ prefers $c\succ a \succ b\succ d$. $j$ prefers $d\succ a \succ b \succ c$. Initially $i$ votes for $a$ and $j$ votes for $d$. 

There are no opportunity moves since $c$ does not dominate $a$ for $i$ (if $i$ moves, $b$ may win with additional two votes). Thus the first step is by $j$ who deserts $d$ and goes to $a$. Now $b$ is no longer a possible winner, and $a$ is dominated (for $i$) by $c$. Thus in the next step $i$ does an opportunity move and the score of the winner decreases.

\bibliography{plurality}
\end{document}